\documentclass[a4paper,10pt]{article}
\usepackage[pdftex]{graphicx}
\PassOptionsToPackage{hyphens}{url}\usepackage{hyperref}
\usepackage{amsmath,amsfonts,amssymb}
\usepackage{fullpage}
\usepackage{authblk}
\usepackage{setspace}
\usepackage{caption}
\usepackage{subcaption}
\usepackage{comment}
\usepackage{booktabs}
\usepackage{tabularx}
\usepackage{adjustbox}
\usepackage[dvipsnames]{xcolor}
\usepackage{float}

\usepackage[explicit]{titlesec}
\usepackage{sidecap}
\usepackage{pbox}
\usepackage[superscript]{cite}

\usepackage{enumitem}
\usepackage{multirow}
\usepackage{longtable}
\usepackage{makecell}
\usepackage{xcolor}
\date{\today}
\title{\bf{Emergence and structure of decentralised trade networks around dark web marketplaces}}

\author[1,2]{Matthieu Nadini}
\author[1]{Alberto Bracci}
\author[3]{Abeer ElBahrawy}
\author[3]{Philip Gradwell}
\author[4]{Alexander Teytelboym}
\author[1,2,5,*]{Andrea Baronchelli}

\affil[1]{\small City, University of London, Department of Mathematics, London EC1V 0HB, UK}
\affil[2]{\small The Alan Turing Institute, British Library, 96 Euston Road, London NW12DB, UK}
\affil[3]{\small Chainalysis Inc, NY, USA}
\affil[4]{\small Department of Economics and INET Oxford, University of Oxford, Oxford OX1 3UQ, UK}
\affil[5]{\small UCL Centre for Blockchain Technologies, University College London, UK}

\affil[*]{\small Corresponding author: Andrea.Baronchelli.1@city.ac.uk}

\begin{document}

\maketitle

Dark web marketplaces (DWMs) are online platforms that facilitate illicit trade among millions of users generating billions of dollars in annual revenue. Recently, two interview-based studies have suggested that DWMs may also promote the emergence of direct user-to-user (U2U) trading relationships. Here, we quantify the scale of, and thoroughly investigate, U2U trading around DWMs by analysing 31 million Bitcoin transactions among users of 40 DWMs between June 2011 and Jan 2021. We find that half of the DWM users trade through U2U pairs generating a total trading volume greater than DWMs themselves. We then show that hundreds of thousands of DWM users form stable trading pairs that are persistent over time. Users in stable pairs are typically the ones with the largest trading volume on DWMs. Then, we show that new U2U pairs often form while both users are active on the same DWM, suggesting the marketplace may serve as a catalyst for new direct trading relationships. Finally, we reveal that stable U2U pairs tend to survive DWM closures and that they were not affected by COVID-19, indicating that their trading activity is resilient to external shocks. Our work unveils sophisticated patterns of trade emerging in the dark web and highlights the importance of investigating user behaviour beyond the immediate buyer-seller network on a single marketplace. 

\medskip


\medskip

\small{\textbf{ORCID.} Matthieu Nadini: 0000-0003-4542-7481; Alberto Bracci: 0000-0002-9506-5645; Abeer ElBahrawy: 0000-0002-4717-891X; Alexander Teytelboym: 0000-0002-6570-1903; Andrea Baronchelli: 0000-0002-0255-0829.}
\newpage
\section*{Introduction}
\label{Introduction}

Since the launch of Silk Road, the first modern dark web marketplace (DWM), in 2011~\cite{christin2013traveling} millions of buyers and sellers have traded in the dark web. DWMs have became popular because their users can anonymously access them through ad-hoc browsers, such as The Onion Router (Tor)~\cite{dingledine2004tor}, and trade goods using cryptocurrencies, such as Bitcoin~\cite{nakamoto2008Bitcoin}. They offer a variety of illicit goods including drugs, firearms, credit cards dumps, and fake IDs~\cite{GwernDarkNets}. DWMs could represent a threat for the regular economy and public health. For instance, during the COVID-19 pandemic, DWMs sold COVID-19 related goods (e.g., masks and COVID-19 tests) that were in shortage in regulated marketplaces as well as unapproved vaccines and fake treatments~\cite{broadhurstavailability, bracci2020covid, bracci2021covid}. Law enforcement agencies have therefore targeted DWMs and users trading on them, performing dozens of arrests and seizing millions of US dollars worth of Bitcoin~\cite{Operation_Onymous, FBIAlphabay, 
BillionFedsSilkRoad}. Despite police raids and unexpected closures, DWM trading volume has been steadily increasing and exceeded \$1.5 billion for the first time in 2020~\cite{Chainalysis_crypto_crime_report_2021}.

DWM users display complex trading patterns within the marketplace environment.
For example, users migrate to alternative DWMs when a DWM that they trade on close~\cite{elbahrawy2020collective, hiramoto2020measuring}. Such migration of users is aided by communication via online forums and chats on the dark web~\cite{buxton2015rise, maddox2016constructive}. However, little is known about how DWM users trade and transact \textit{outside} the DWMs. On the one hand, some recent works have shown that a significant number of DWM users  trade drugs and other illicit goods using social media platforms, such as Facebook, Telegram, and Reddit~\cite{oksanen2020illicit, DarknetLive_telegram, sung2021prevalence, childs2021beyond, kwon2021dark}. Moreover, several qualitative, interview-based studies have shown that DWM users form direct trading relationships with other users, starting user-to-user (U2U) pairs that bypass the intermediary role of DWMs~\cite{barratt2016if, munksgaard2020and}. Past research has also found that sellers on regulated online marketplaces and social medial platforms may decide to use intermediaries, such Facebook groups or Instagram, to find new customers, and may start direct U2U trading with potential buyers~\cite{bakken2019sellers}. In this paper, we look closely at patterns of U2U trading relationships among DWM users.

\begin{figure}[H]
  \centering
 \begin{subfigure}[H]{0.45\textwidth}
         \centering
         \vspace{1cm}
         \includegraphics[width=0.7\textwidth]{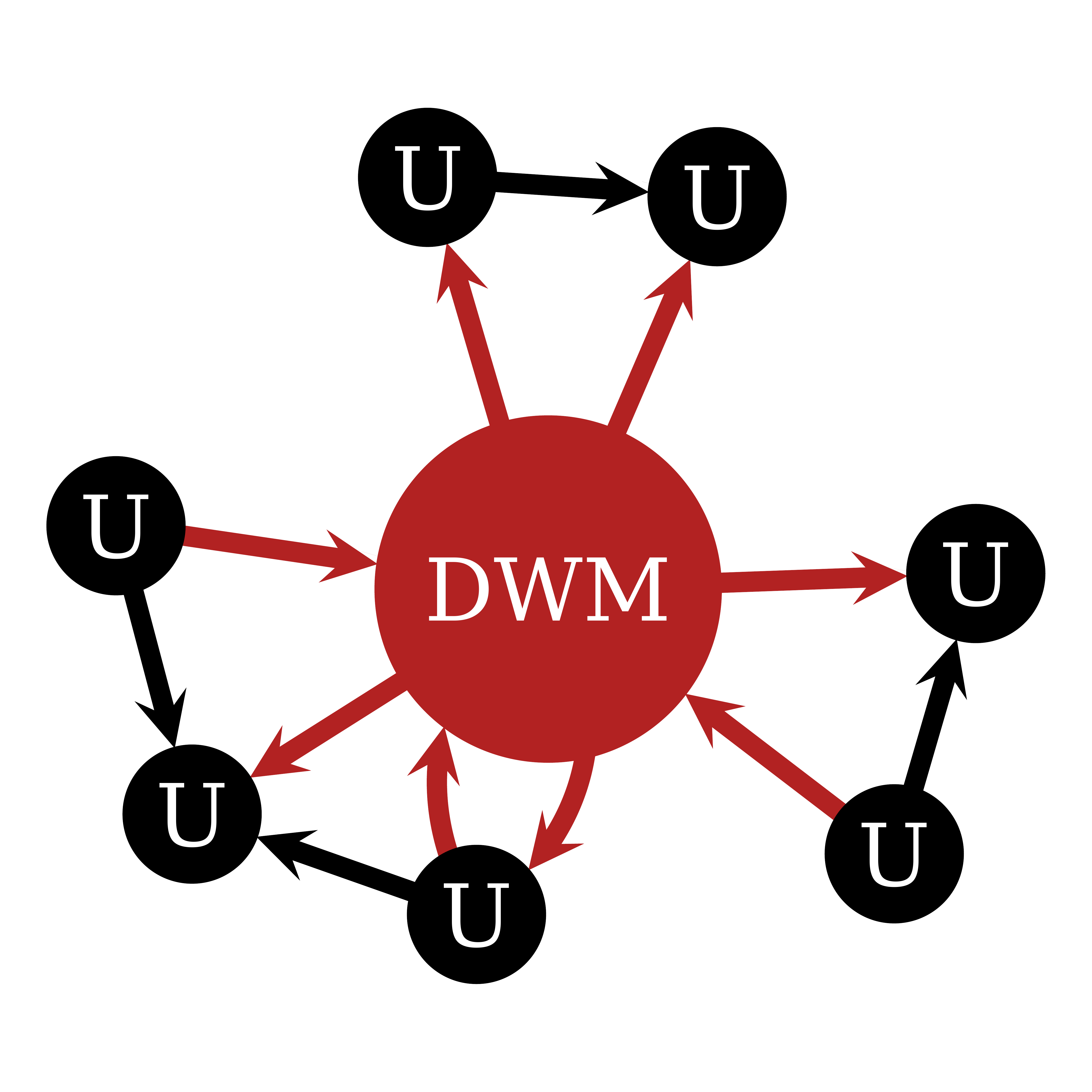}
         \vspace{1cm}
     \caption*{(a) Marketplace ego network}
 \end{subfigure}
  \begin{subfigure}[H]{0.45\textwidth}
         \centering
         \includegraphics[width=\textwidth]{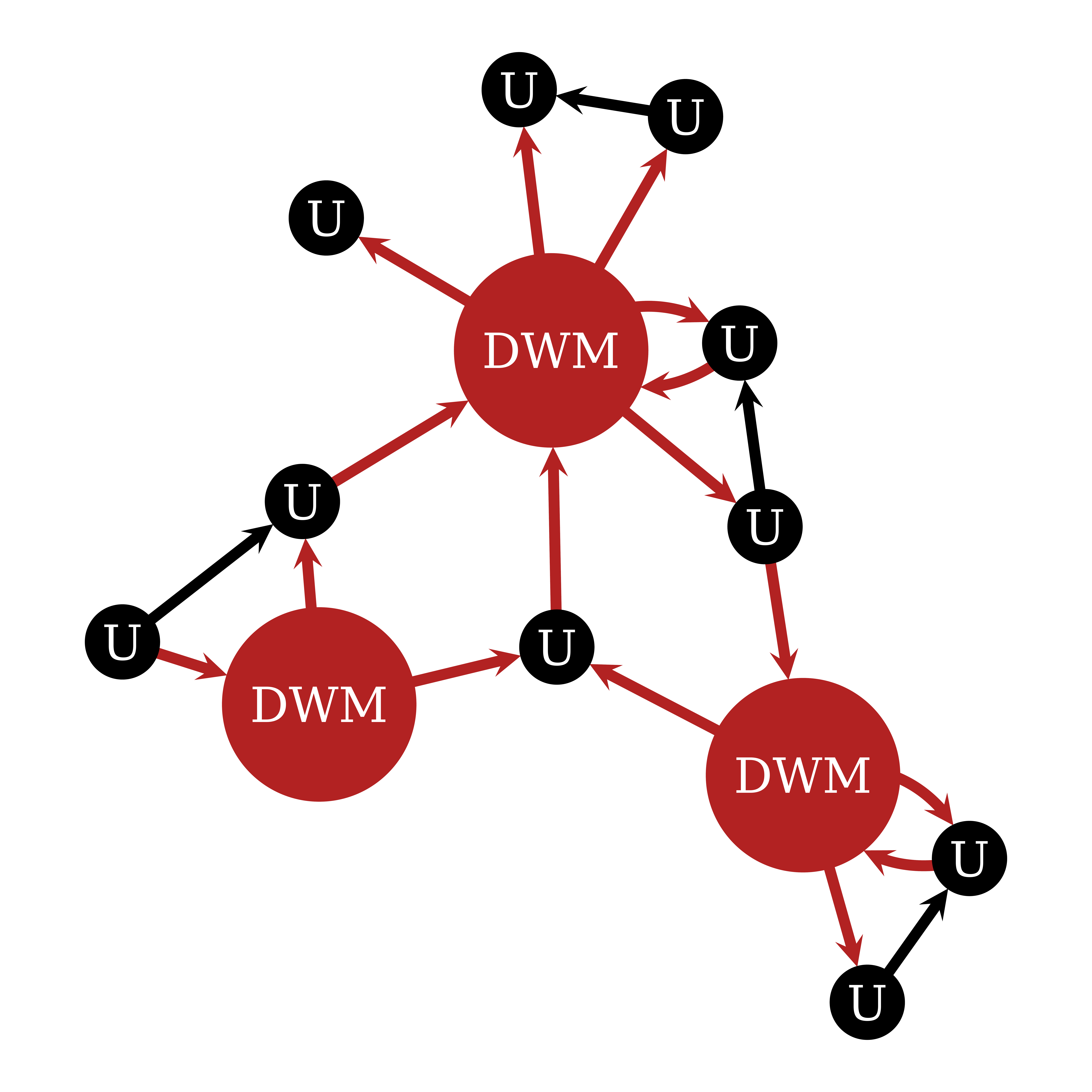}
     \caption*{(b) Full network}
 \end{subfigure}
  \caption{\textbf{Ego and full networks.} (a) Schematic representation of an ego network surrounding a dark web marketplace (``DWM'', in red). The DWM interacts with its users (``U'', in black), which make user-to-user (U2U) pairs, represented with arrows and their respective users. (b) Multiple ego networks may be aggregated to form the full network.}
  \label{DWM_ego_full_networks}
\end{figure}

The starting point for this paper is identifying U2U networks around DWMs. We analyse 40 DWMs for a 10-year time period spanning from June 18, 2011 to January 31, 2021. Our dataset covers all major DWMs that have ever existed, as identified by the European Monitoring Centre, Europol, the World Health Organization, and independent researchers~\cite{european2017drugs, world2019world, gwern_live_markets}. Our analysis focuses on Bitcoin -- the most popular cryptocurrency on DWMs~\cite{lee2019cybercriminal, foley2019sex} as well as in the regulated economy~\cite{baur2015cryptocurrencies, saiedi2020global}. We focus on two kinds of transactions, occurring (i) between the user and a DWM and (ii) between two users of the same DWM. The result is 40 distinct marketplace ego networks containing  user-DWM and U2U transactions, whose typical structure is depicted in Figure~\ref{DWM_ego_full_networks}(a). In each network, links are directed and the arrows point at the receiver of Bitcoin. Since users often migrate from one DWM to another~\cite{elbahrawy2020collective} and become users of multiple DWMs, the 40 ego networks are not isolated, and can be combined to form one full network, as shown in Figure~\ref{DWM_ego_full_networks}(b).

Previous analyses of U2U trading relationships around DWMs include only two studies~\cite{barratt2016if, munksgaard2020and} based on  unstructured~\cite{barratt2016if} or semi-structured~\cite{munksgaard2020and} interviews of  17 users of Silk Road and 13 DWMs sellers, respectively. Here, we dramatically extend previous work by exploring the collective emergence and structure of U2U pairs. First, we observe that the U2U network, formed by all transactions between pairs of users, has a larger trading volume than DWMs themselves. We then identify stable U2U trading relationships, which represent a subset of persistent pairs in our dataset~\cite{nadini2020detecting, nadini2020reconstructing} forming the \emph{backbone} of the U2U network. We find that 137,667 (i.e., 1.7\% out of 7.85 million total) pairs are stable, generating a total trading volume of \$1.5 billion (i.e., 5\% out of \$30 billion total volume). We then explore the behaviour of users forming stable U2U pairs. We reveal that stable U2U pairs play a crucial role for marketplaces by spending significantly more time and generating far greater transaction volume with DWMs than other users. By analysing the temporal evolution of stable pairs, we unveil that DWMs acted as meeting points for 37,192 (out of around 16 million users), whose trading volume is estimated to be \$417 million. Importantly, these newly formed pairs persist in time and transact for several months even after the closure of the DWM that spurred their formation. Finally, we observe that COVID-19 only had a temporary impact on the evolution of stable U2U pairs, which continued to increase their trading volume throughout 2020.

\section*{Results}
\label{Results}

\subsection*{Large number of U2U transactions}

\paragraph{Ego networks.}

\begin{figure}[H]
  \centering
  \includegraphics[width=16cm]{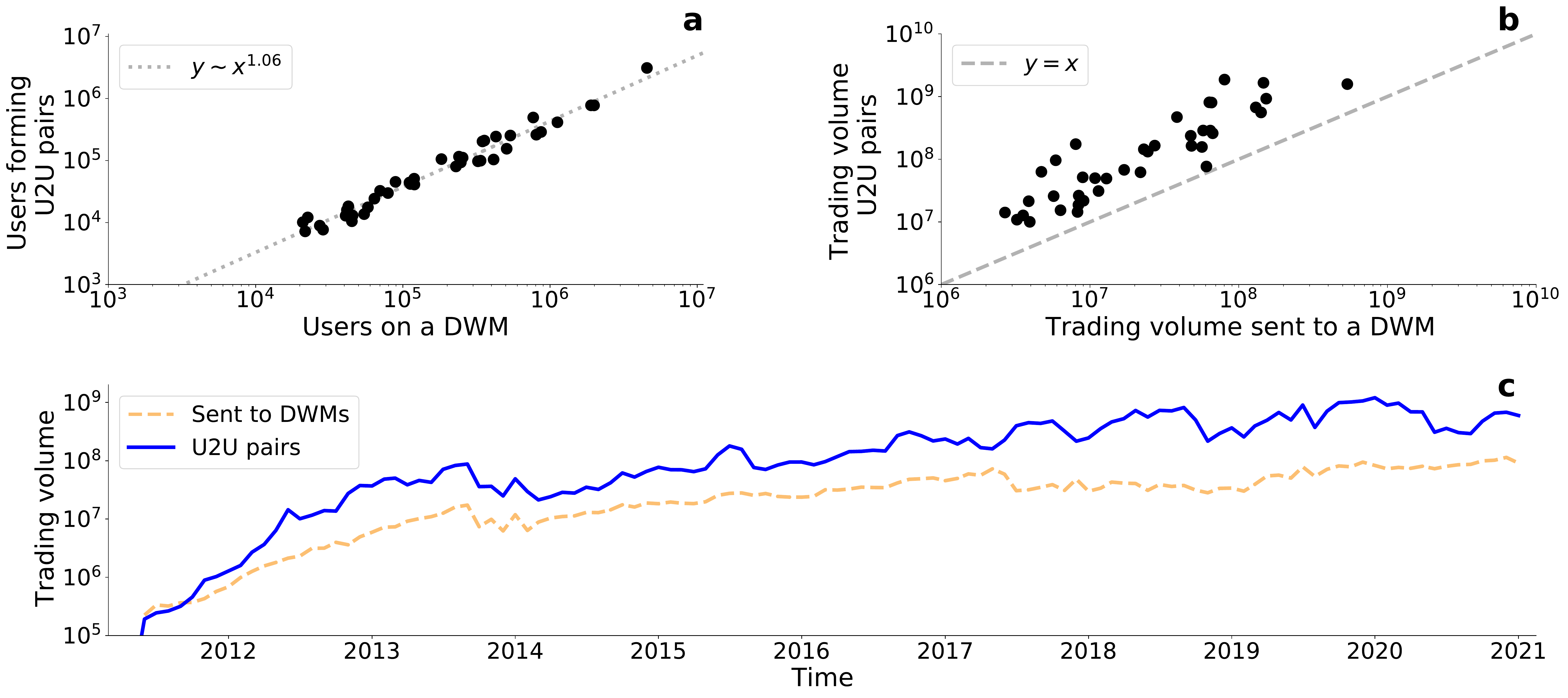}
  \caption{\textbf{User-DWM and U2U transactions.} (a) Total number of users interacting with a DWM against the total number of them forming U2U transactions. The dotted line corresponds to the result of a fitted power law function. (b) Trading volume in dollars sent to a DWM compared with the total trading volume in its surrounding U2U transactions. The dashed line is the bisector and allows to easily compare the two trading volumes. (c) Total monthly trading volume sent to all DWMs and exchanged in all unique U2U pairs. We do not include the trading volume received from DWMs because it is equivalent as the volume sent to DWMs.}
  \label{Importance_U2U_transactions}
\end{figure}

We start our analysis by measuring the extent of the U2U network around each DWM. The percentages of users forming U2U pairs vary across DWMs, with a median value of 38\% (min 23\%, max 68\%). The variance in the percentages of users with U2U pairs is shown by Figure~\ref{Importance_U2U_transactions}(a), which shows that the number of users with U2U pairs obeys an almost linear relationship with the number of users interacting with a DWM, having an exponent equal to 1.06 and $R^2 = 0.969$. The total trading volume users sent to the marketplace is obviously equivalent to the one they receive from it (two-sided Wilcoxon test~\cite{wilcoxon1992individual}: $W=330$, $p=0.282$). Importantly, the total trading volume users sent to a DWM (and consequently the one that they receive from it) is always less than the one exchanged through U2U transactions, as shown in Figure~\ref{Importance_U2U_transactions}(b).

\paragraph{Full network.}

\begin{figure}[H]
  \centering
  \includegraphics[width=16cm]{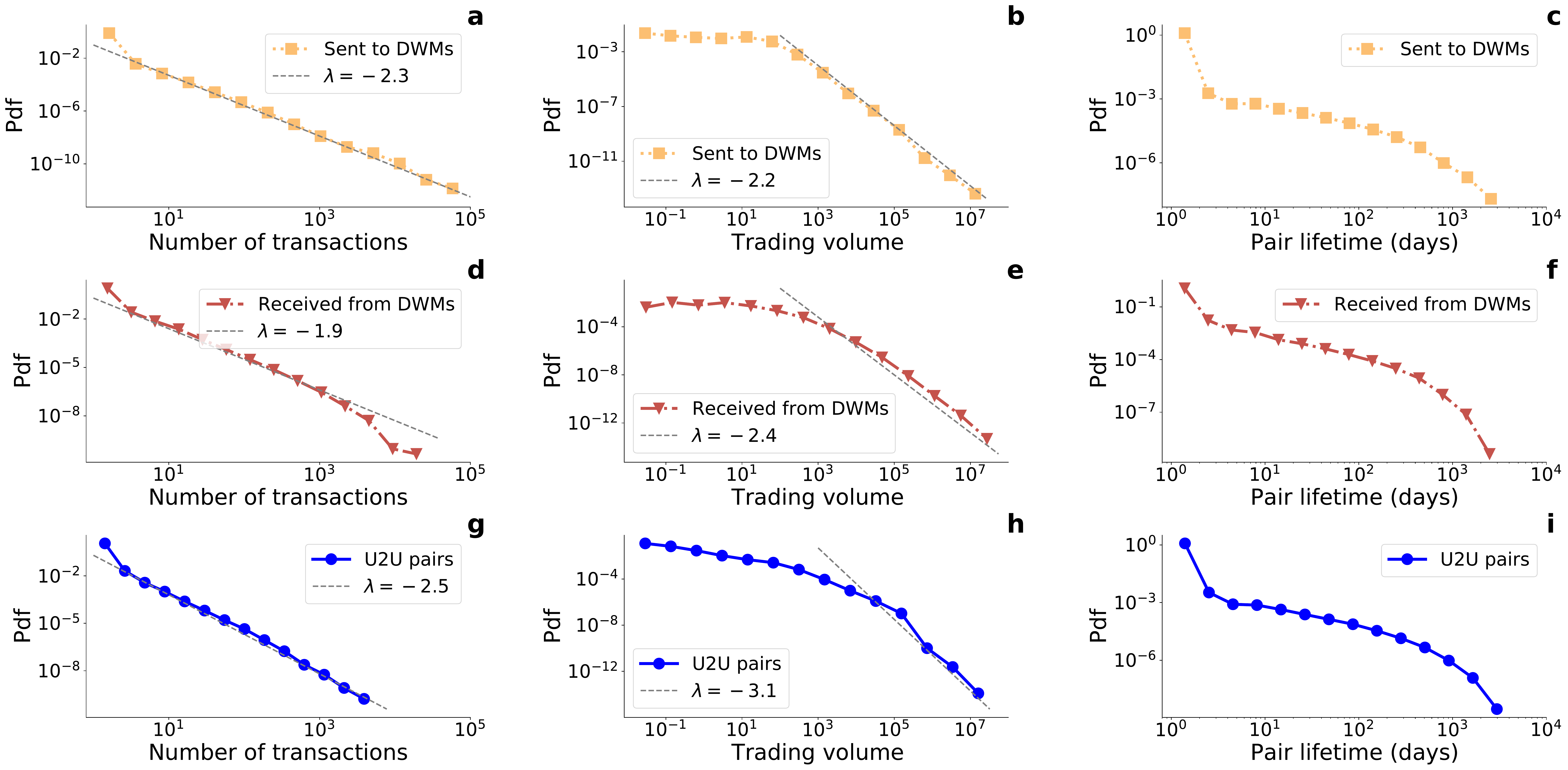}
  \caption{\textbf{Key distributions of the full network.} Probability distribution function (pdf) about the number of transactions of each pair of entities (a)-(d)-(g), their trading volume (b)-(g)-(h), and their lifetime computed as time difference between their last and first transaction (c)-(f)-(i).}
  \label{Distributions_pairs_dataset}
\end{figure}

Similar results hold for the full network, confirming that the formation of U2U pairs is a pervasive phenomenon around DWMs. The total trading volume users sent to DWMs is \$3.8 billion, received from DWMs \$3.7 billion, while the volume exchanged through U2U pairs reaches \$30 billion. 
In Figure~\ref{Distributions_pairs_dataset}, we illustrate the number of transactions, trading volume, and lifespan of U2U pairs. In all cases we observe familiar fat-tailed distributions.

We then consider the temporal evolution of transactions. We look at the trading volume over time in Figure~\ref{Importance_U2U_transactions}(c), where we observe that U2U transactions have consistently involved greater monthly volume than the volume sent to all DWMs since 2011. This underlines the economic importance of U2U transactions in the Bitcoin ecosystem relative to DWMs.

\subsection*{Behaviour of the U2U network}

\begin{figure}[H]
  \centering
  \includegraphics[width=16cm]{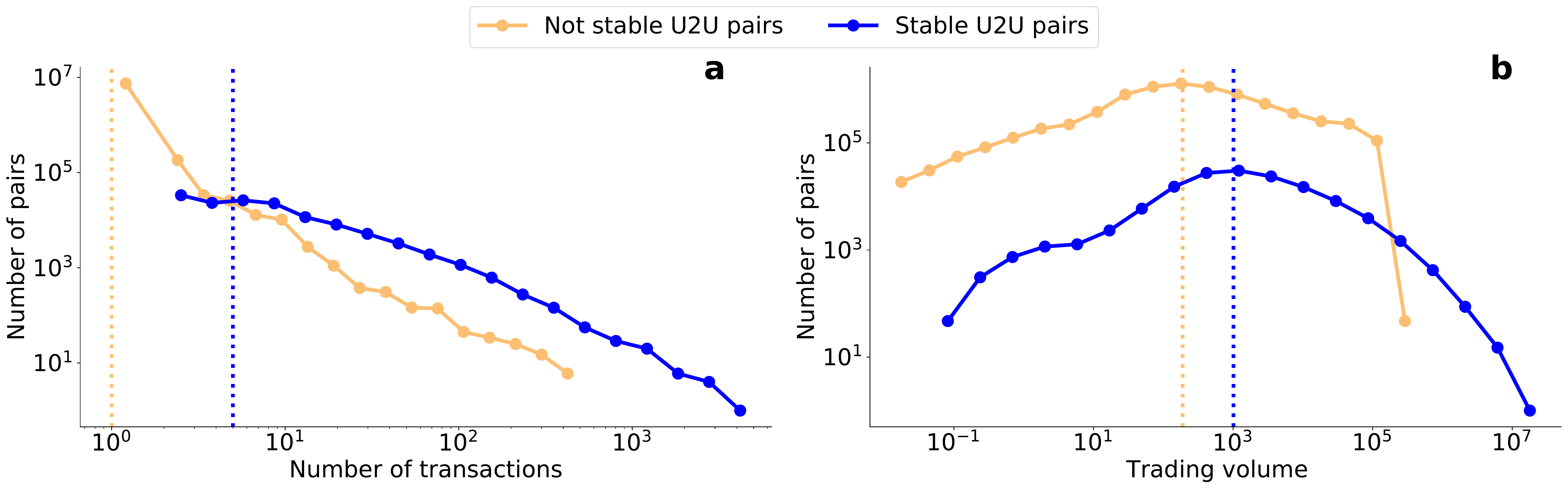}
  \caption{\textbf{Statistics of U2U pairs.} (a) Number of stable and non-stable U2U pairs with a given number of transactions. (b) Number of stable and non-stable U2U pairs with a given trading volume. Vertical lines represent median values of the respective distributions.} 
  \label{Number_pairs_stable_not}
\end{figure}

Henceforth, we are going to analyse users by focusing on the following groups: users who do not form stable U2U pairs; users who form stable U2U pairs, of which there are users who met outside DWMs and users who met inside DWMs (see the nomenclature in Table~\ref{Nomenclature}). We start by focusing our attention on identifying stable U2U pairs, i.e., persistent pairs of the U2U network. To this end, we use the evolving activity-driven model~\cite{nadini2020detecting} to extract them in a statistically-principled way (see Methods). We find 137,667 stable U2U pairs formed by 106,648 users and generating a trading volume equal to \$1.5 billion. Stable pairs produce five times more transactions per pair than non-stable pairs (two-sided Mann-Whitney-U test~\cite{mann1947test}: MNU$=4,58 \cdot 10^9$, $p<0.0001$) corresponding to a 5.34 times larger trading volume (MNU$= 317 \cdot 10^9$, $p<0.0001$), see  Figure~\ref{Number_pairs_stable_not}. Stable pairs, despite representing less than 2\% of the total number of U2U pairs, generate a disproportionate amount of trading volume.

\begin{figure}[H]
  \centering
  \includegraphics[width=12cm]{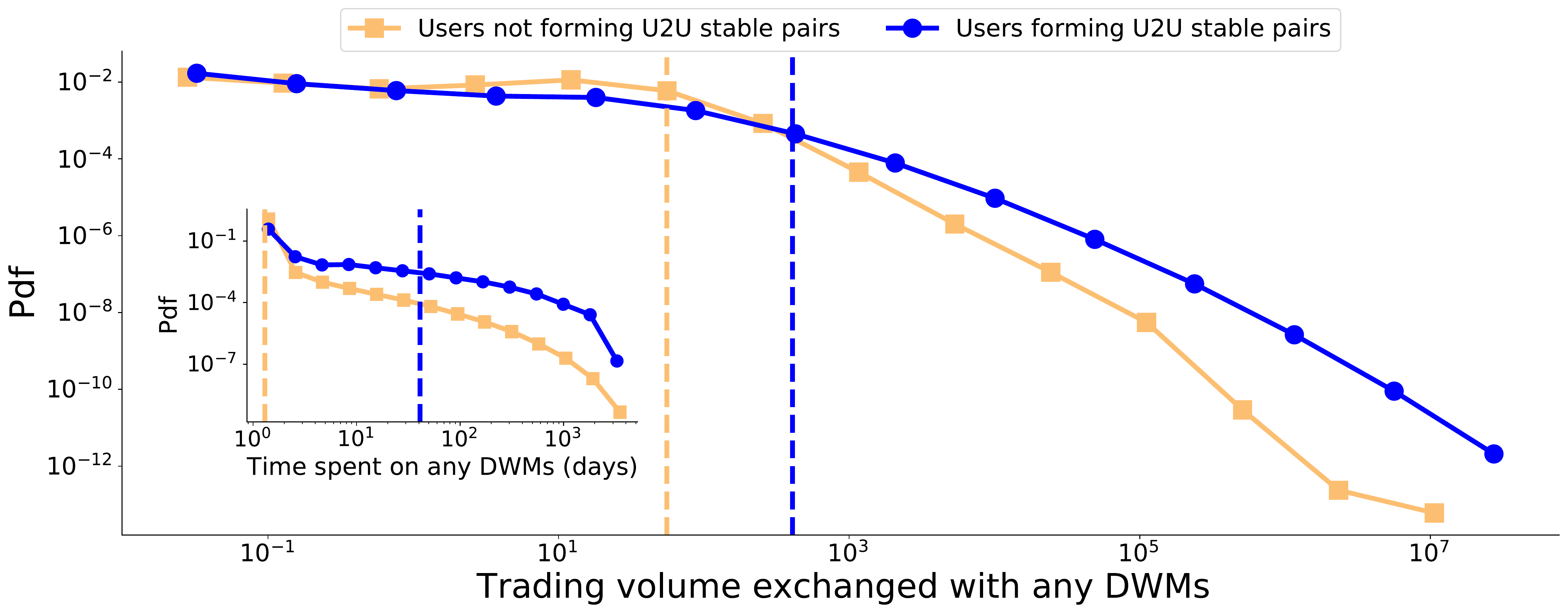}
  \caption{\textbf{Role of users forming stable U2U pairs.} (Main) PDFs of trading volume that users exchange with any DWMs. (Inset) PDFs of time spent by users on any DWMs. These distributions are explored for each of the 40 DWMs under consideration in Figure~\ref{Role_users_in_DWMs_time_spent} and~\ref{Role_users_trading_volume}, respectively. Vertical lines represent median values of the respective distributions.}
  \label{Role_users}
\end{figure} 

\begin{figure}[H]
  \centering
  \includegraphics[width=16cm]{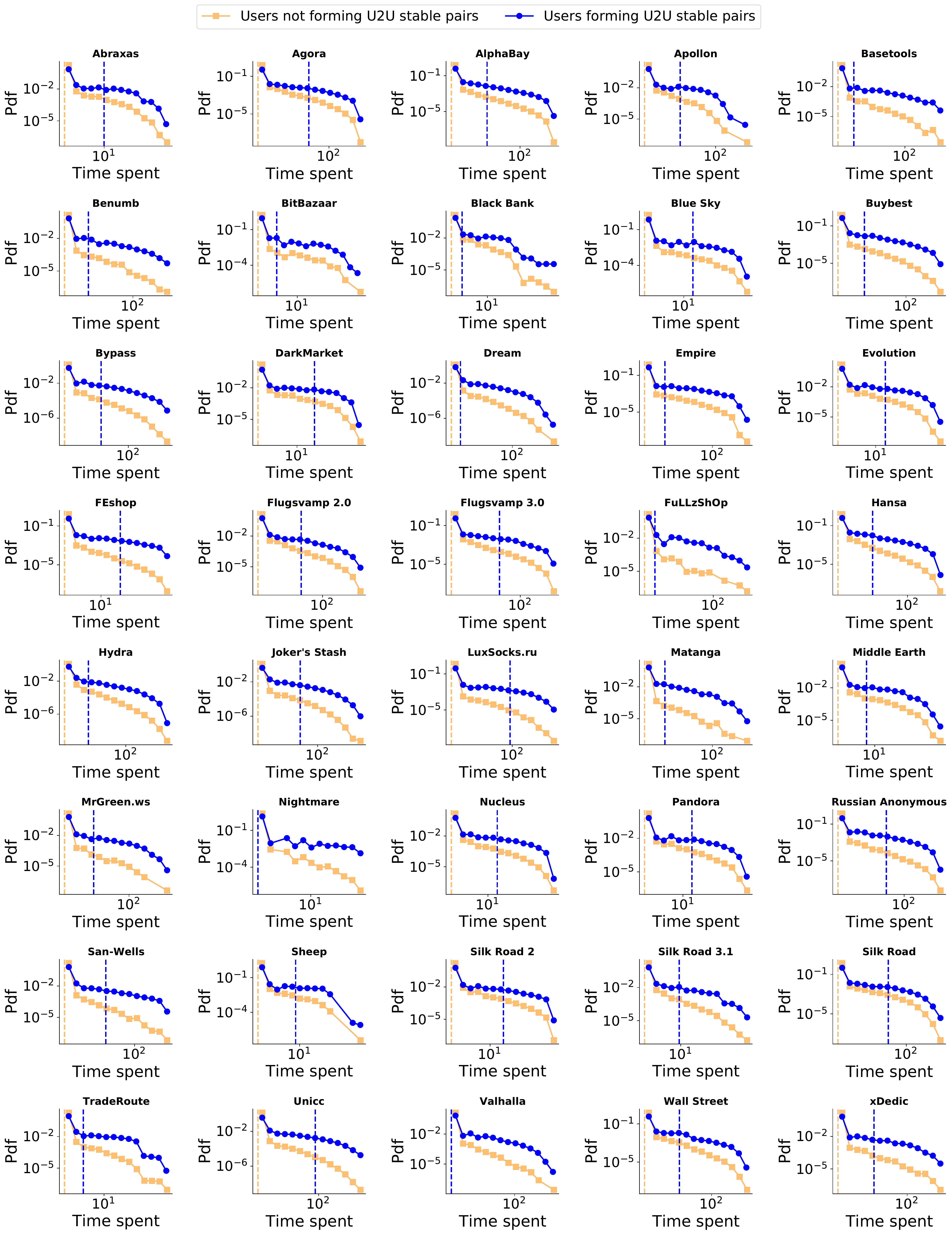}
  \caption{\textbf{Evaluation of the time users spent on a DWM.} It extends Figure~\ref{Role_users}(inset) in the main text by considering each individual DWM. Statistical tests are carried using the two-sided Kolmogorov-Smirnov test and results are available in Table~\ref{Statistical_tests}. Vertical lines represent median values of the respective distributions.}
 \label{Role_users_in_DWMs_time_spent}
\end{figure}

\begin{figure}[H]
  \centering
  \includegraphics[width=16cm]{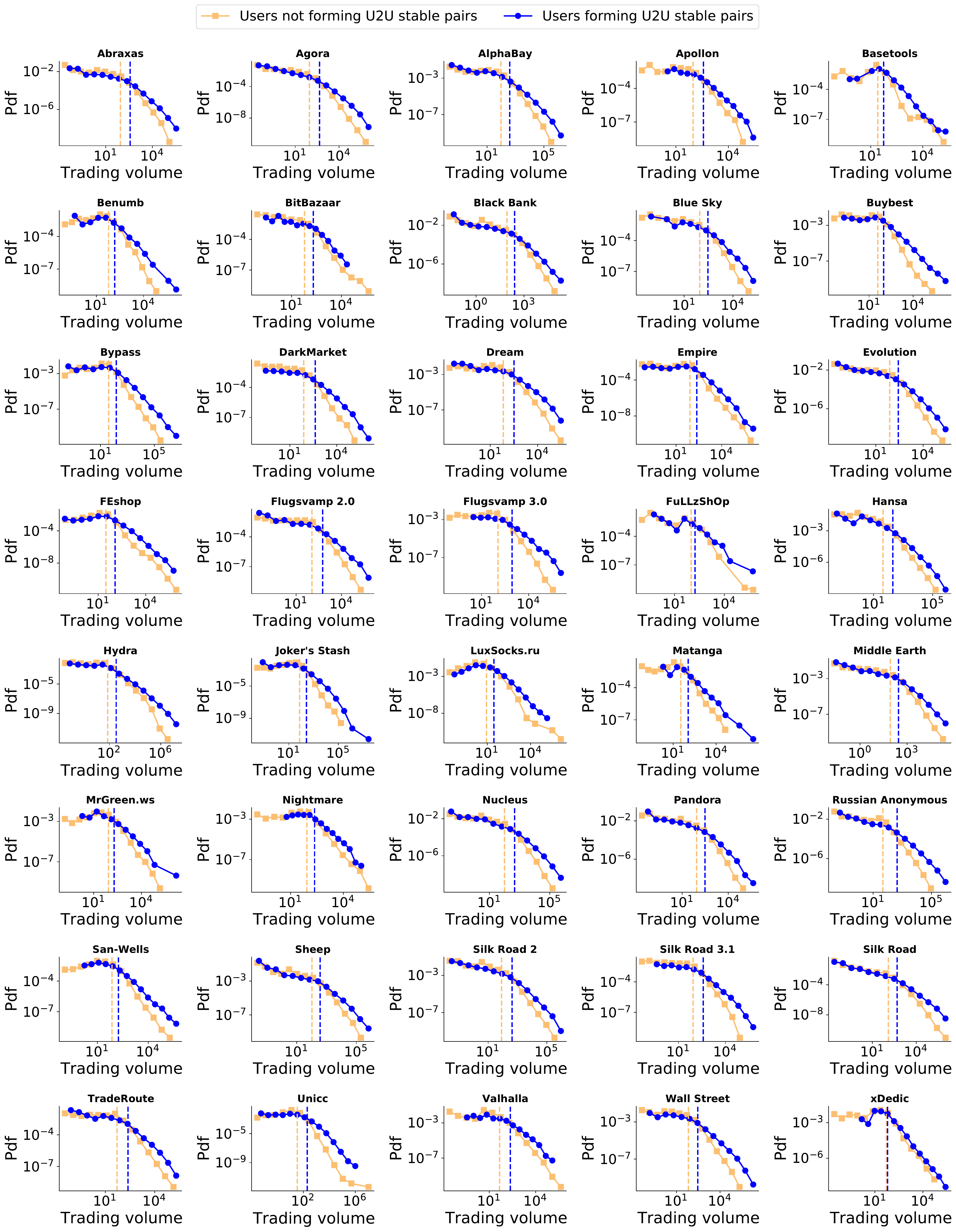}
  \caption{\textbf{Evaluation of the total trading volume users exchange with a DWM.} It extends Figure~\ref{Role_users} in the main text by considering each individual DWM. Statistical tests are carried using the two-sided Kolmogorov-Smirnov test and results are available in Table~\ref{Statistical_tests}. Vertical lines represent median values of the respective distributions.}
  \label{Role_users_trading_volume}
\end{figure}

The high activity of users forming stable U2U pairs is not limited to the U2U network, as they are also the most active in trading with DWMs. 
Users in stable U2U pairs spend a median number of 41 days on DWMs versus a median of only one day for users without stable pairs. 
The two resulting distributions are significantly different (two-sided Kolmogorov-Smirnov test~\cite{massey1951kolmogorov}: KS $= 0.673$, $p<0.0001$), see the inset of Figure~\ref{Role_users}. When we look at the trading volume  with DWMs, we find qualitatively similar results. Users in stable U2U pairs transact a median of \$400 with DWMs, while other users transact only \$56. The two resulting distributions are significantly different (KS $= 0.438$, $p<0.0001$), see Figure~\ref{Role_users}. These results hold not only for full network but for every DWM in our data, see Figure~\ref{Role_users_in_DWMs_time_spent} and~\ref{Role_users_trading_volume}.

\subsection*{U2U network evolution}

\paragraph{Formation of U2U stable pairs.}

\begin{table}[H]
\centering
\begin{tabular}{ccc|ccc|ccc}
\multicolumn{3}{c|}{\thead{Users who met outside the DWM}} & \multicolumn{3}{c|}{\thead{Users who met outside the DWM}} & \multicolumn{3}{c}{\thead{Users who met inside the DWM}} \\ 
\begin{subfigure}[H]{0.08\textwidth}
         \centering
         \includegraphics[width=\textwidth]{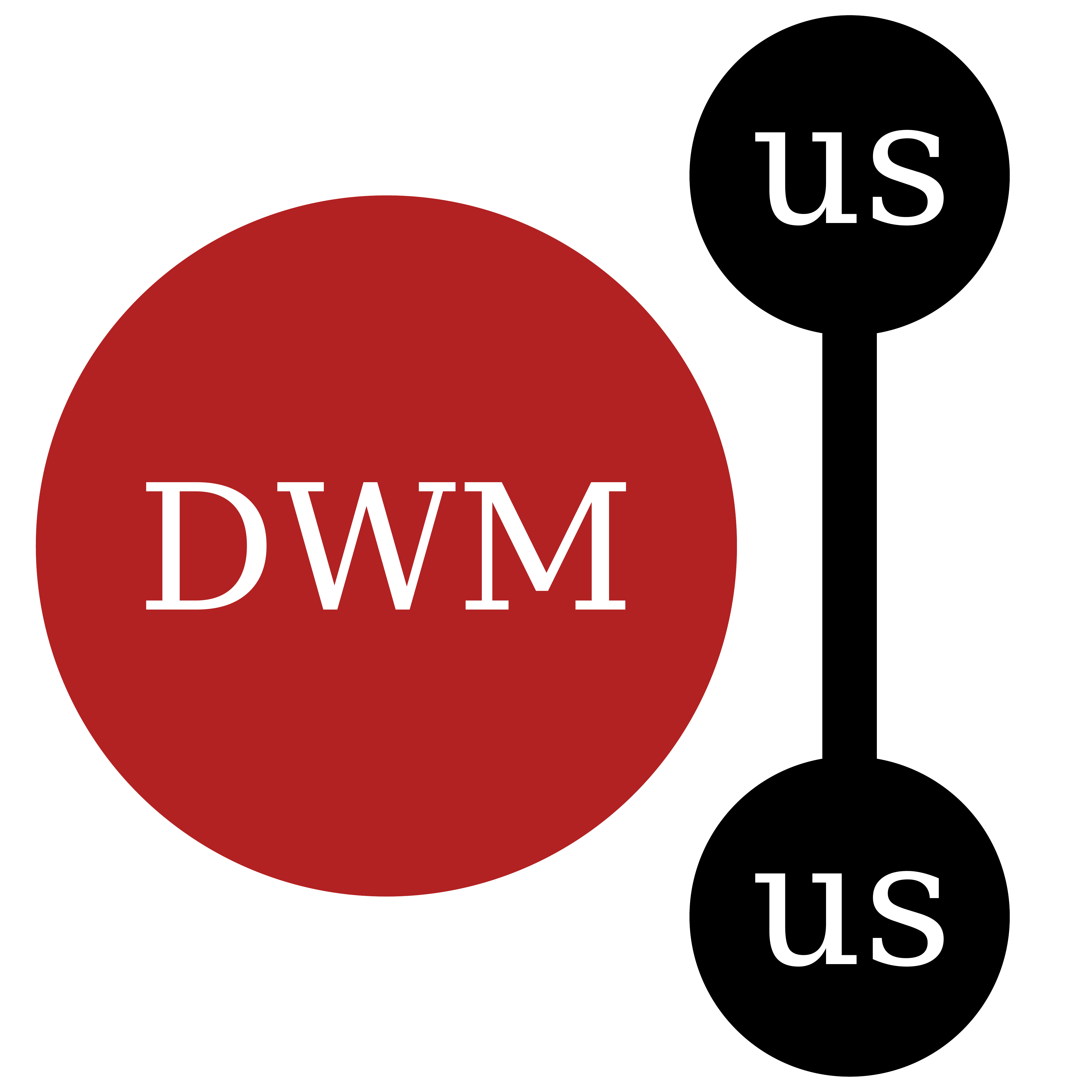}
         \caption*{\thead{$t_1$}}
\end{subfigure}
&
\begin{subfigure}[H]{0.08\textwidth}
         \centering
         \includegraphics[width=\textwidth]{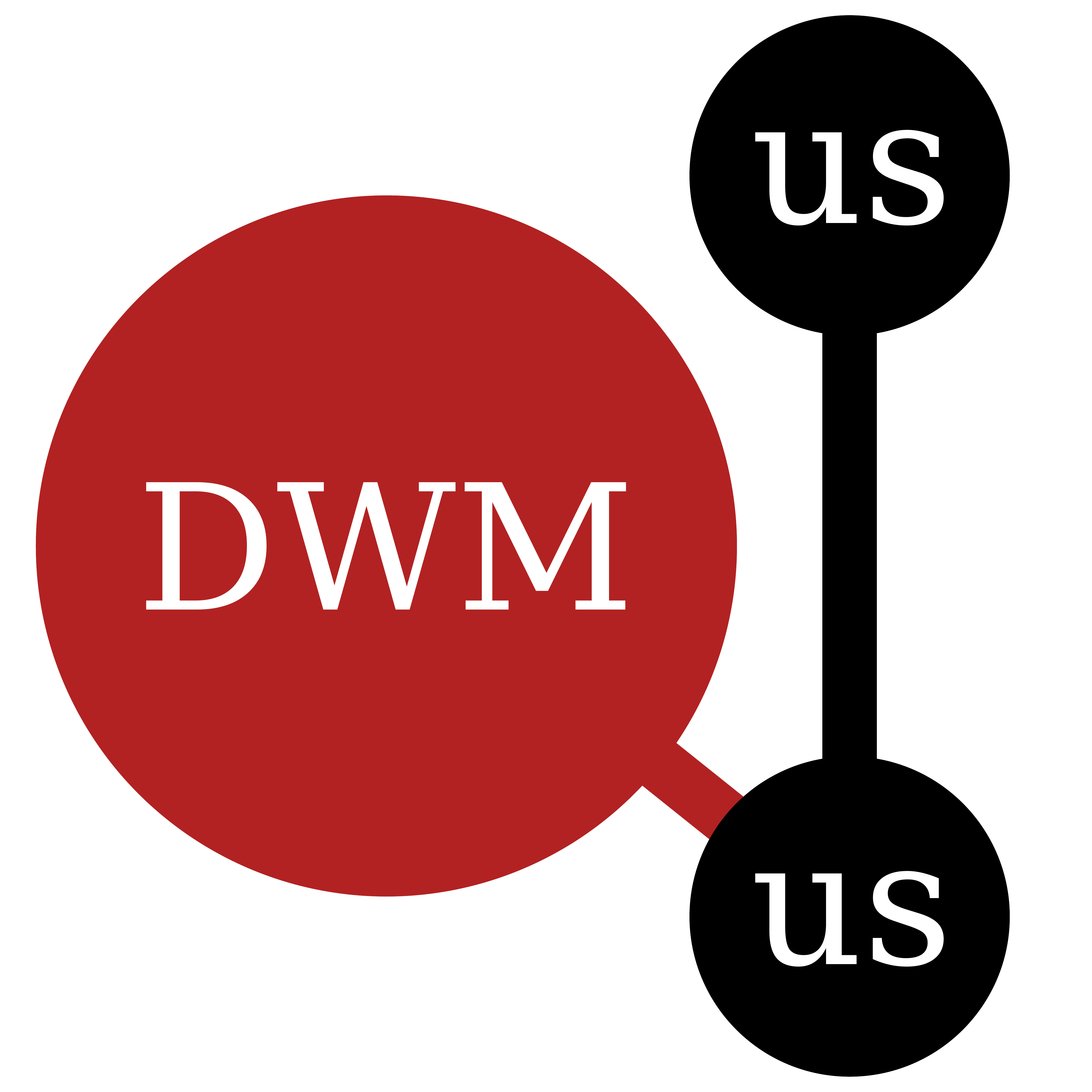}
         \caption*{\thead{$t_2$}}
\end{subfigure}
&
\begin{subfigure}[H]{0.08\textwidth}
         \centering
         \includegraphics[width=\textwidth]{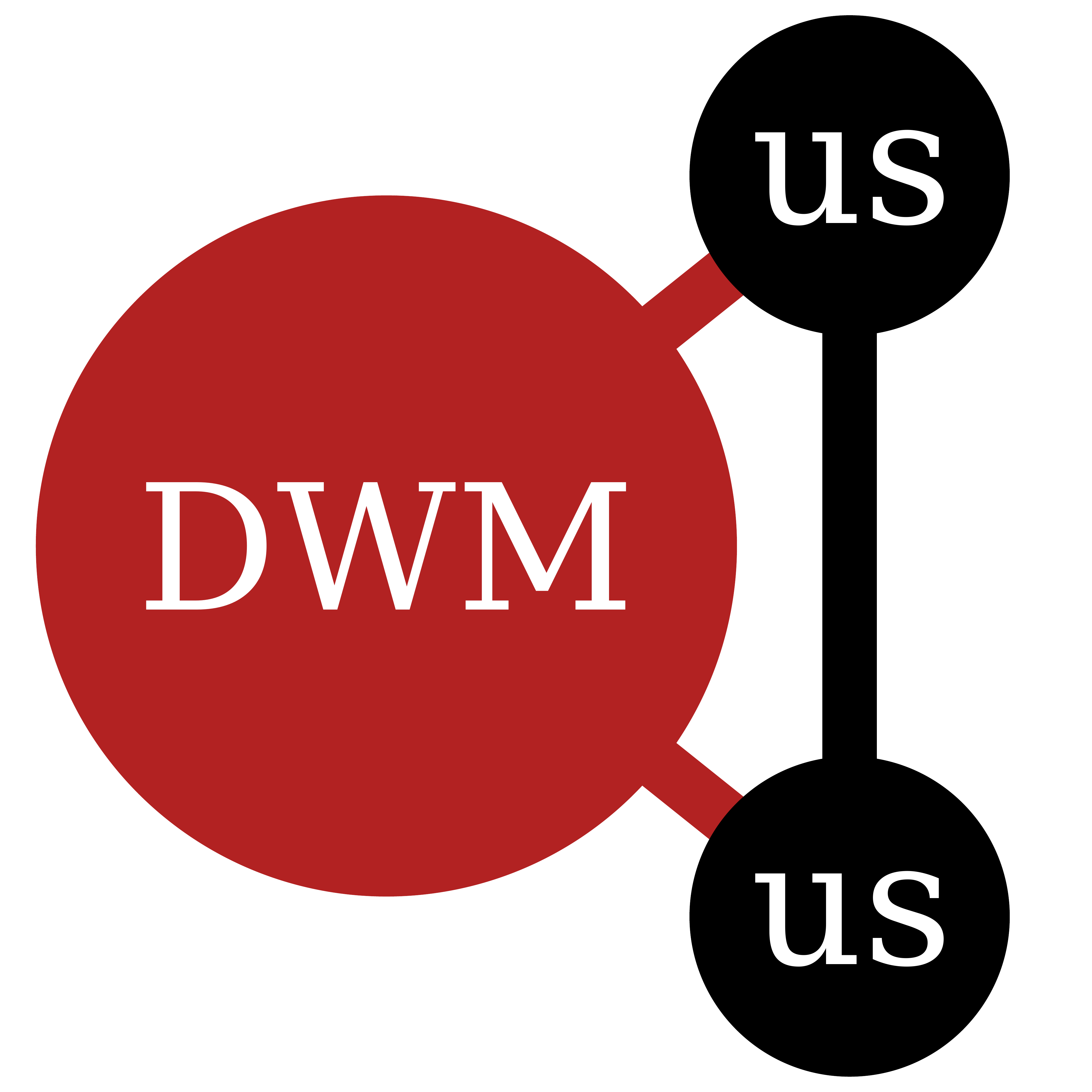}
         \caption*{\thead{$t_3$}}
\end{subfigure} &
 \begin{subfigure}[H]{0.08\textwidth}
         \centering
         \includegraphics[width=\textwidth]{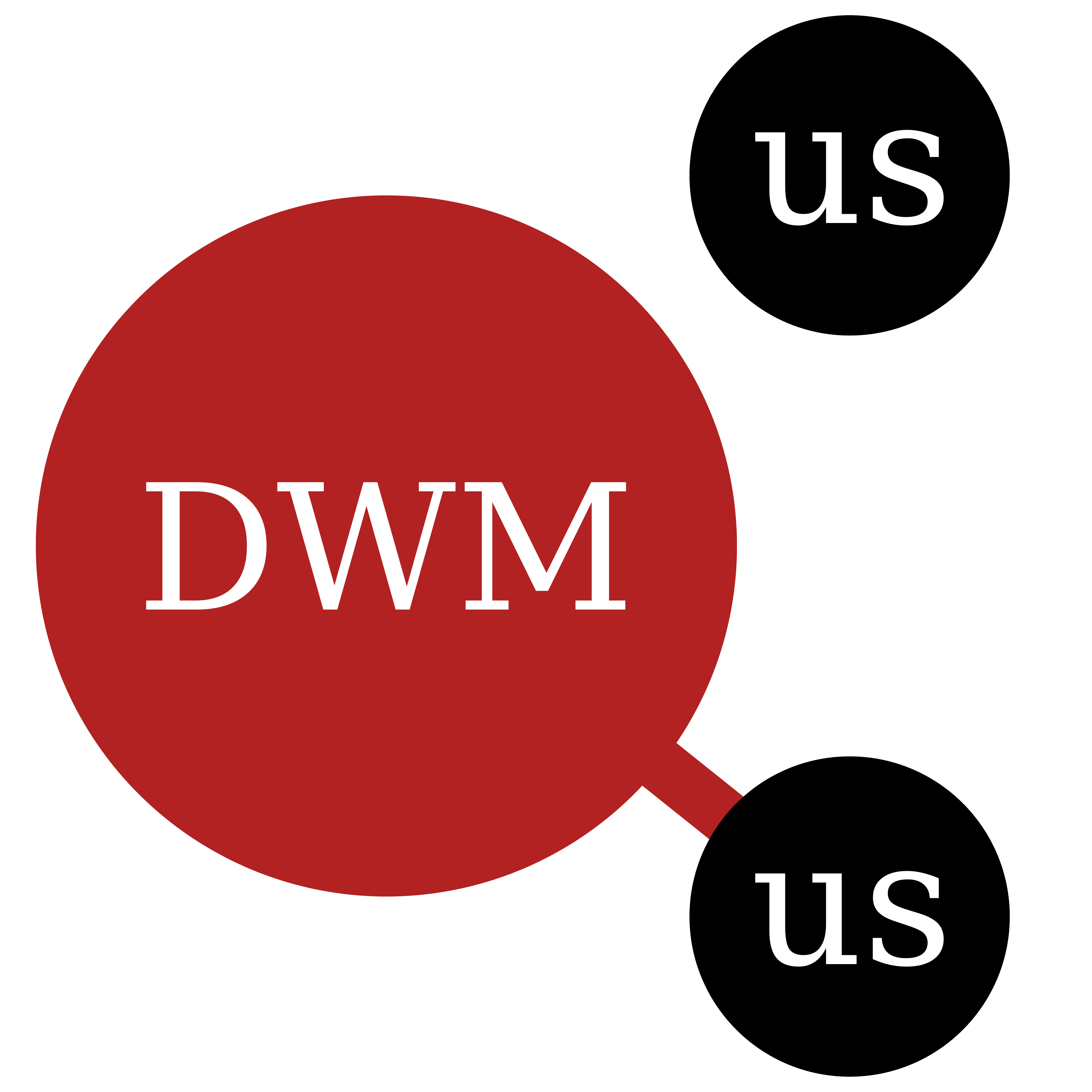}
         \caption*{\thead{$t_1$}}
\end{subfigure}
&
\begin{subfigure}[H]{0.08\textwidth}
         \centering
         \includegraphics[width=\textwidth]{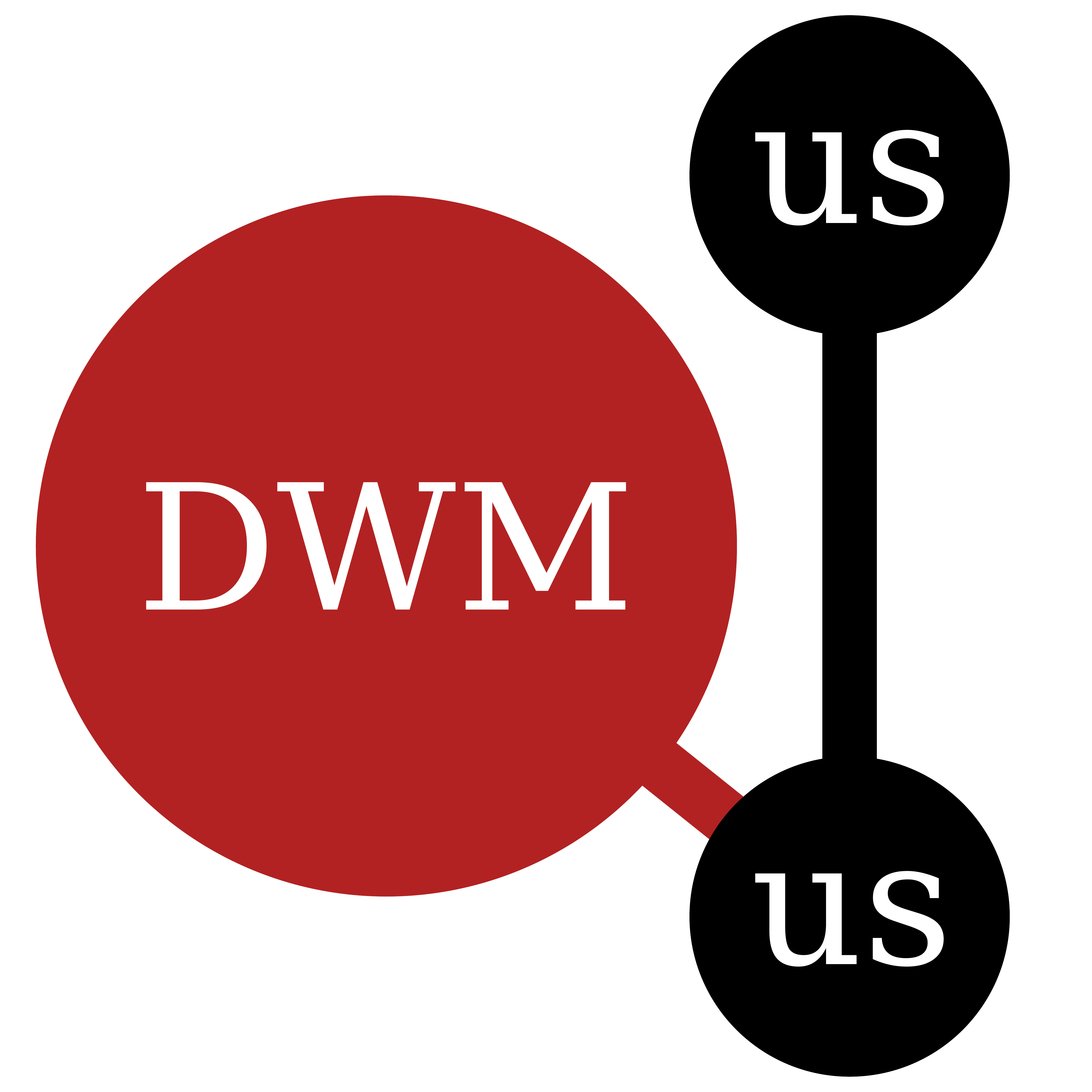}
         \caption*{\thead{$t_2$}}
\end{subfigure}
&
\begin{subfigure}[H]{0.08\textwidth}
         \centering
         \includegraphics[width=\textwidth]{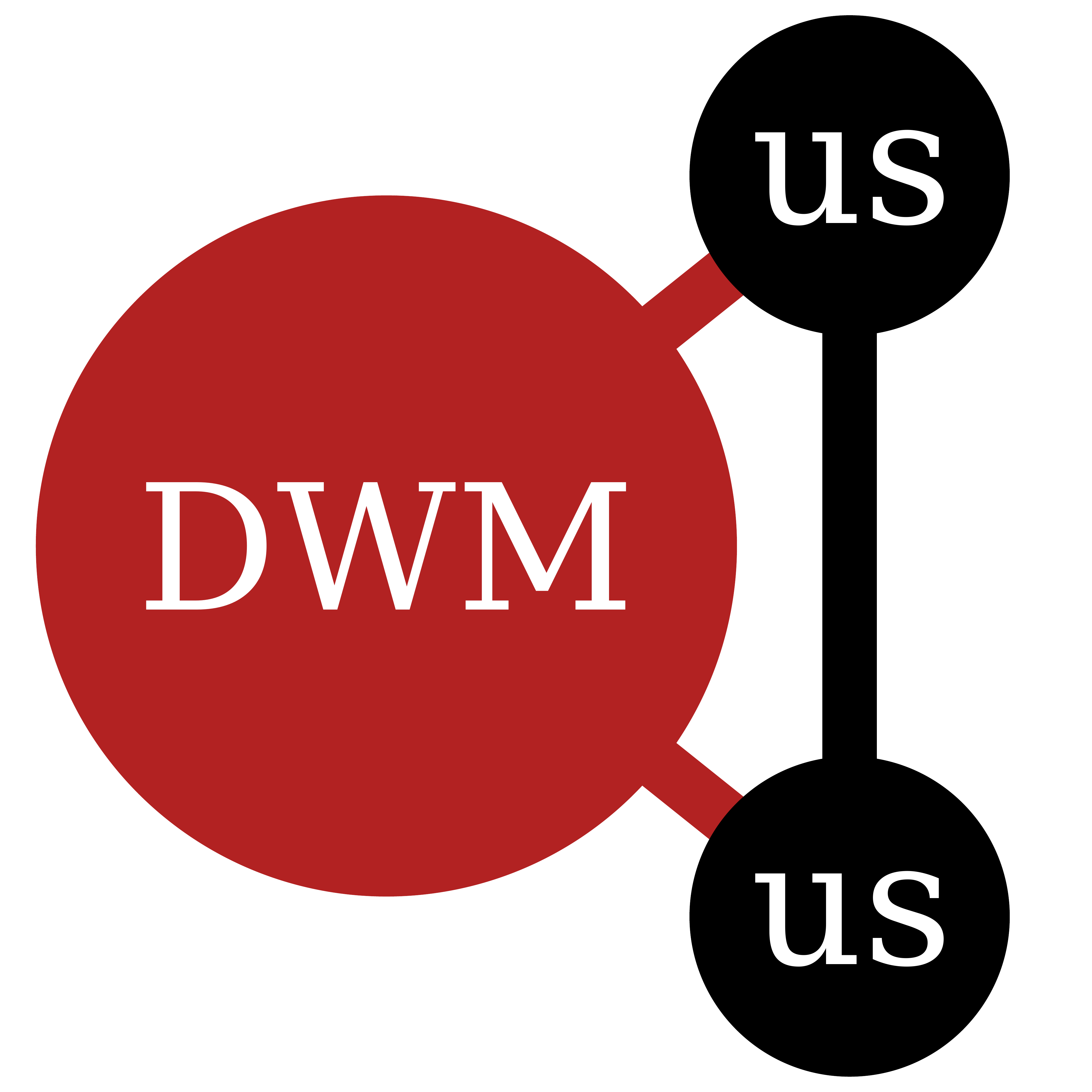}
         \caption*{\thead{$t_3$}}
\end{subfigure} 
&
\begin{subfigure}[H]{0.08\textwidth}
         \centering
         \includegraphics[width=\textwidth]{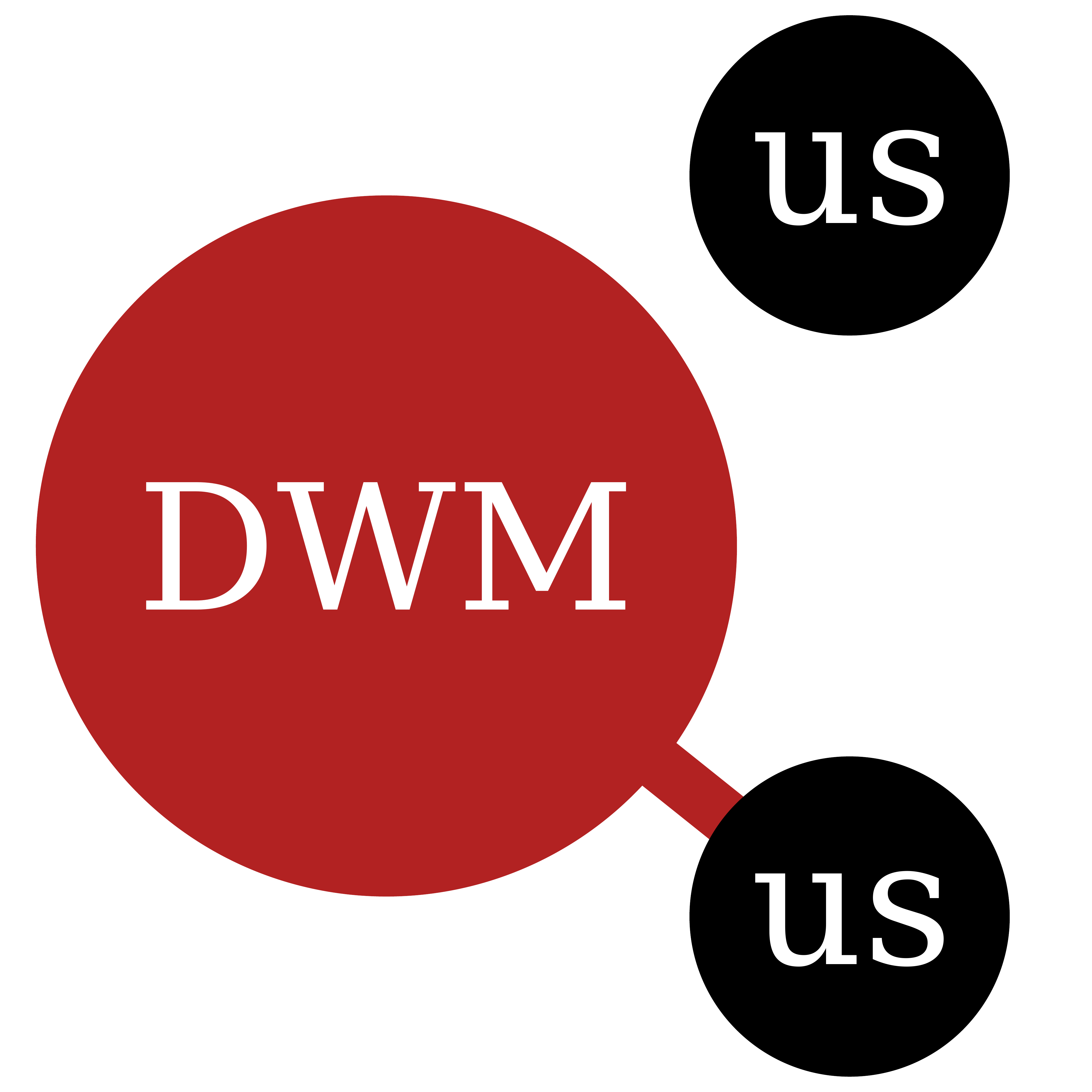}
         \caption*{\thead{$t_1$}}
\end{subfigure}
&
\begin{subfigure}[H]{0.08\textwidth}
         \centering
         \includegraphics[width=\textwidth]{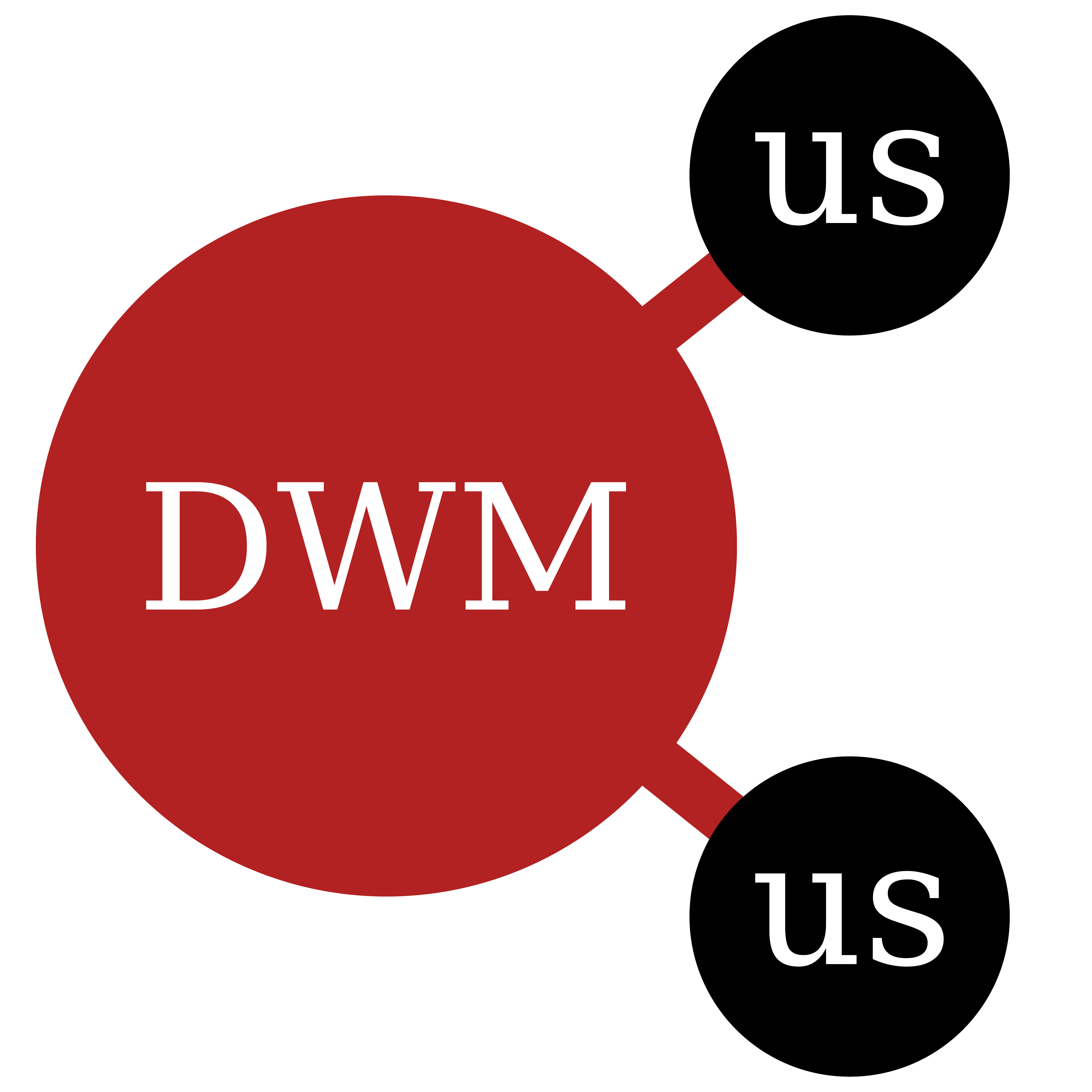}
         \caption*{\thead{$t_2$}}
\end{subfigure}
&
\begin{subfigure}[H]{0.08\textwidth}
         \centering
         \includegraphics[width=\textwidth]{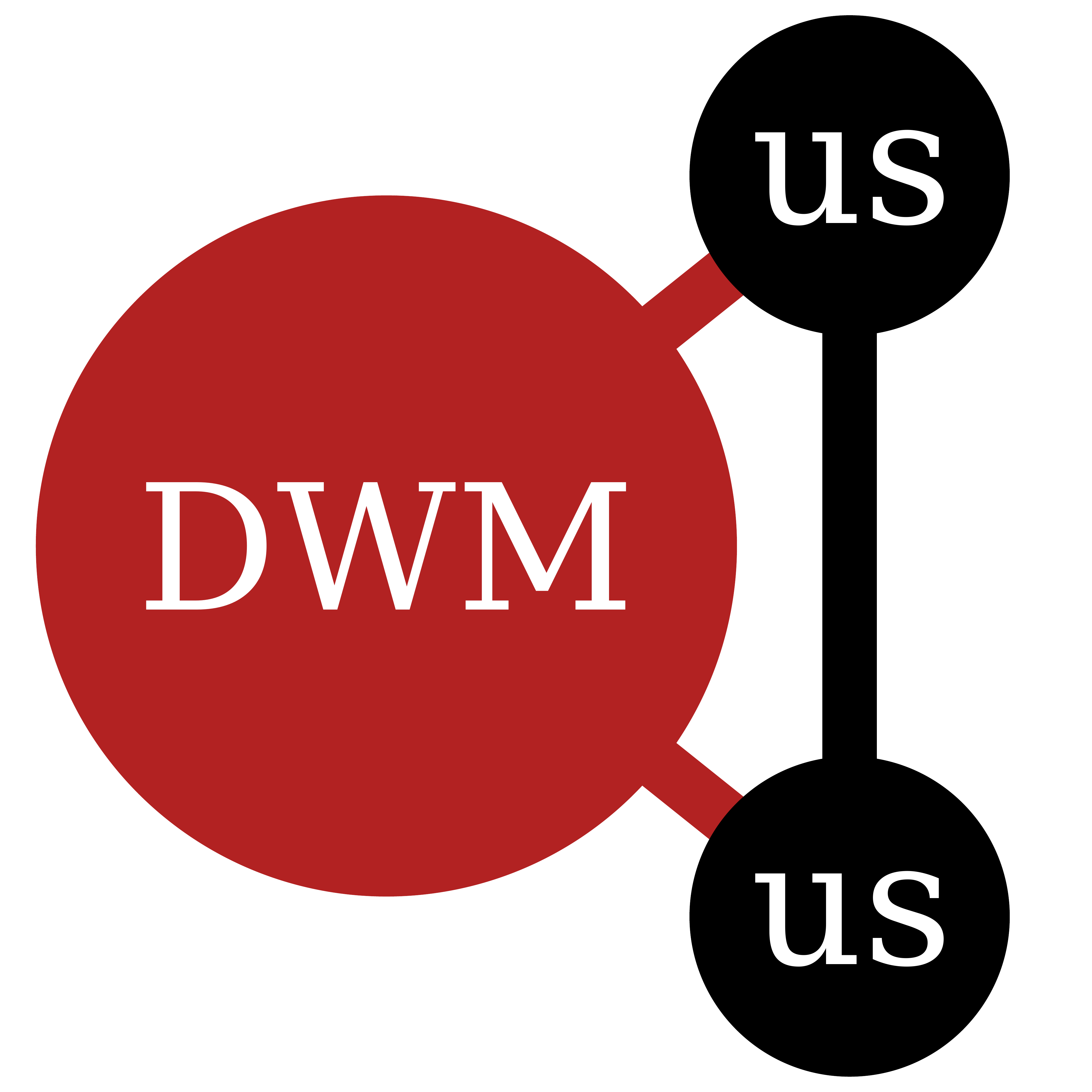}
         \caption*{\thead{$t_3$}}
\end{subfigure}
\end{tabular}
\caption{\textbf{Formation mechanism of stable U2U pairs.} We compare the time in which the first transaction between a pair of users occur with the time in which these users interact with the same DWM. Each row in the figure indicates a possible temporal sequence, which we classify in two groups: users who met outside the DWM (first two columns) and users who met inside the DWM (last column).}
\label{Motifs_Triads_table11}
\end{table}

\begin{figure}[H]
  \centering
  \includegraphics[width=16cm]{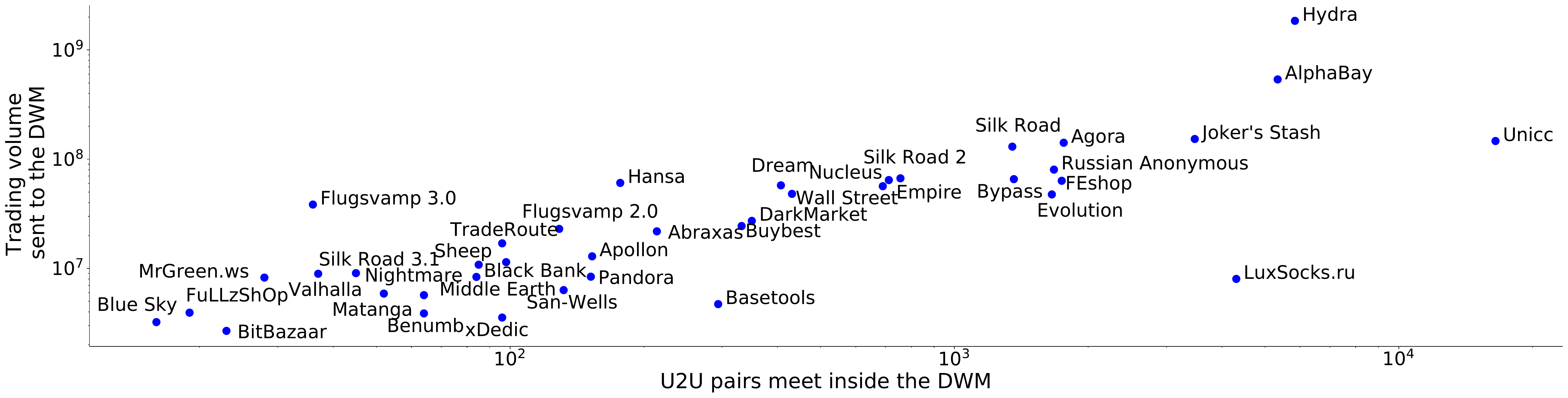}
  \caption{\textbf{DWMs where users meet.} Scatter plot of the number of pairs of users that meet inside each of the 40 DWMs considered versus the total volume sent to the DWM.}
  \label{Where_users_meet}
\end{figure}

Having mapped the behaviour of stable pairs, we now consider their temporal evolution. More specifically, we ask: How do stable pairs form? Do DWMs spur their creation?
One possible hypothesis is that users meet for the first time while active on a DWM, i.e., after they have both traded with that DWM, see Table~\ref{Motifs_Triads_table11} and the nomenclature in Table~\ref{Nomenclature}. This can be considered as a plausible, and conservative, proxy for users who met inside a DWM (see Methods). A total of 37,129 users have met at least one other user inside a DWM.
Their trading volume is about \$417 million, and the percentage of users who met inside a DWM is proportional to the trading volume sent to DWMs (Spearman~\cite{spearman1961proof}: $C=0.805$, $p<0.0001$), see Fig~\ref{Where_users_meet}, meaning that large DWMs are more likely to favour the encounter of users than smaller DWMs. Importantly, users who met inside a DWM transact more than those meeting outside them. In particular, users who met inside a DWM trade a median of \$2,212 between themselves, almost twice the \$1,379 for users meeting outside the DWM (MNU$= 1.863 \cdot 10^9$, $p<0.0001$). Moreover, users who met inside a DWM tend to make transactions significantly longer with median of 61 days than users meeting outside with a median of 50 days (MNU $= 2.099 \cdot 10^9$, $p<0.0001$). 

\begin{figure}[H]
  \centering
  \includegraphics[width=12cm]{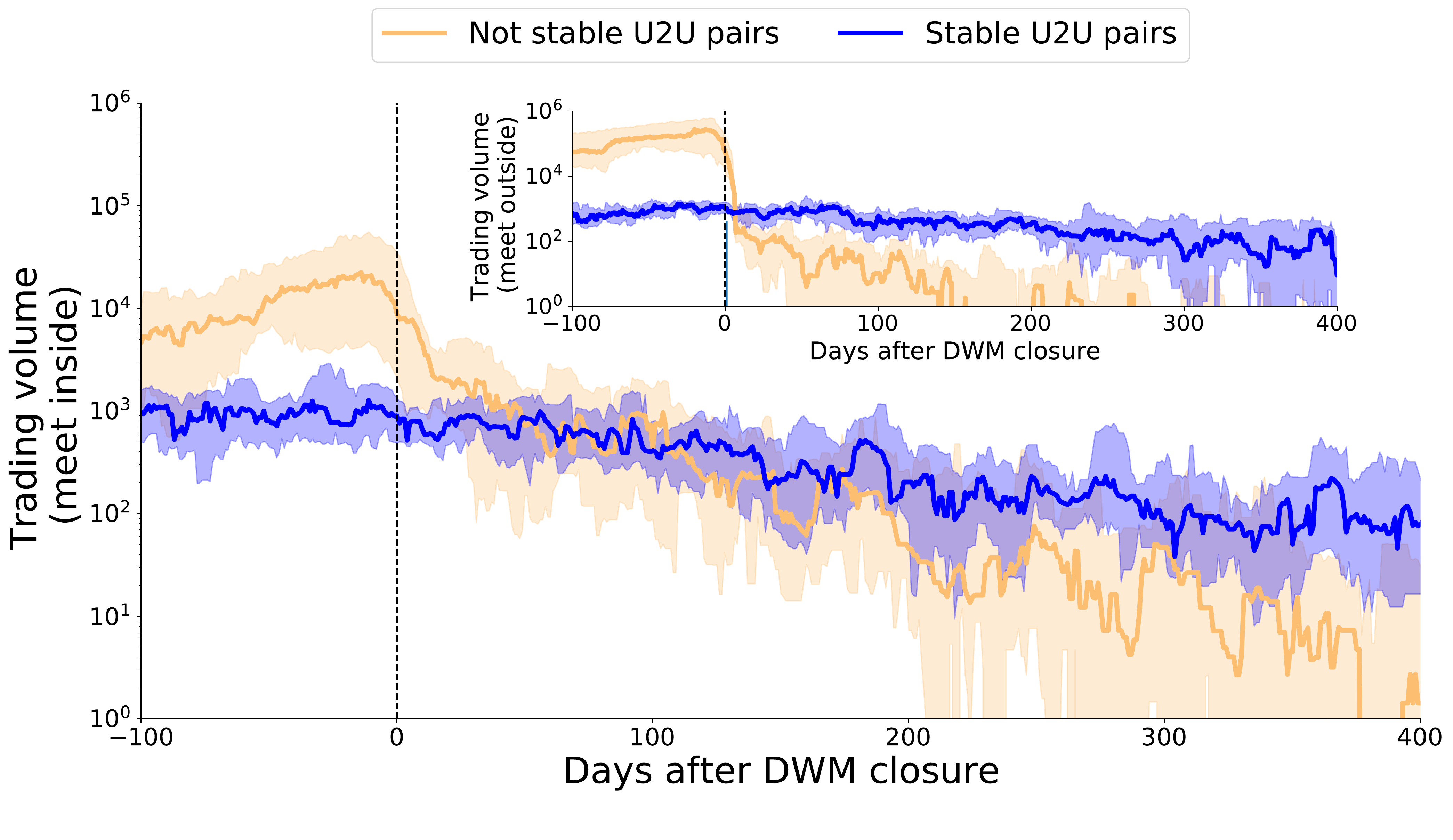}
   \caption{\textbf{Resilience of stable U2U pairs after DWMs closure.} Trading volume of U2U pairs surrounding active DWMs. (Main) U2U pairs meet who met inside aa DWM. (Inset) U2U pairs meet outside them. Curves indicate the median value while bands represent the 95\% confidence interval. Day zero corresponds to the day when the market closed. Negative and positive numbers indicate the days prior and after the closure, respectively. Only the 33 DWMs closed are considered in the analysis.}
  \label{After_DWMs_closure_meet_inside_outside}
\end{figure}

\paragraph{Resilience of U2U stable pairs.}

Thus far, we have shown that users involved in stable trading relationships are also very active on DWMs, where they may meet new trading partners. But are DWMs and the U2U network truly interdependent? In particular, do stable pairs need the DWMs to survive? To answer these questions, we look at market closures, previously investigated to show how active users migrate to other existing DWMs~\cite{elbahrawy2020collective}. Our dataset includes 33 closure events, which we study independently from one another by considering the evolution of the respective 33 marketplace ego networks. We find that non-stable U2U pairs sharply stop interacting immediately after DWM closure therefore their existence is highly sensitive to the presence of the DWM. On the other hand, the trading volume of stable U2U pairs is only marginally affected by the disappearance of the DWM. As a result, while prior to DWM closure non-stable U2U pairs generate an overall trading volume that is 10 times higher than that of stable U2U pairs (since non-stable pairs are far more prevalent), within a few weeks after DWM closure the pattern is reversed: stable U2U pairs generate more trade volume than non-stable U2U pairs. Indeed, trading patterns of stable pairs are not significantly influenced by DWMs closure, see  Figure~\ref{After_DWMs_closure_meet_inside_outside}.

\begin{figure}[H]
  \centering
  \includegraphics[width=16cm]{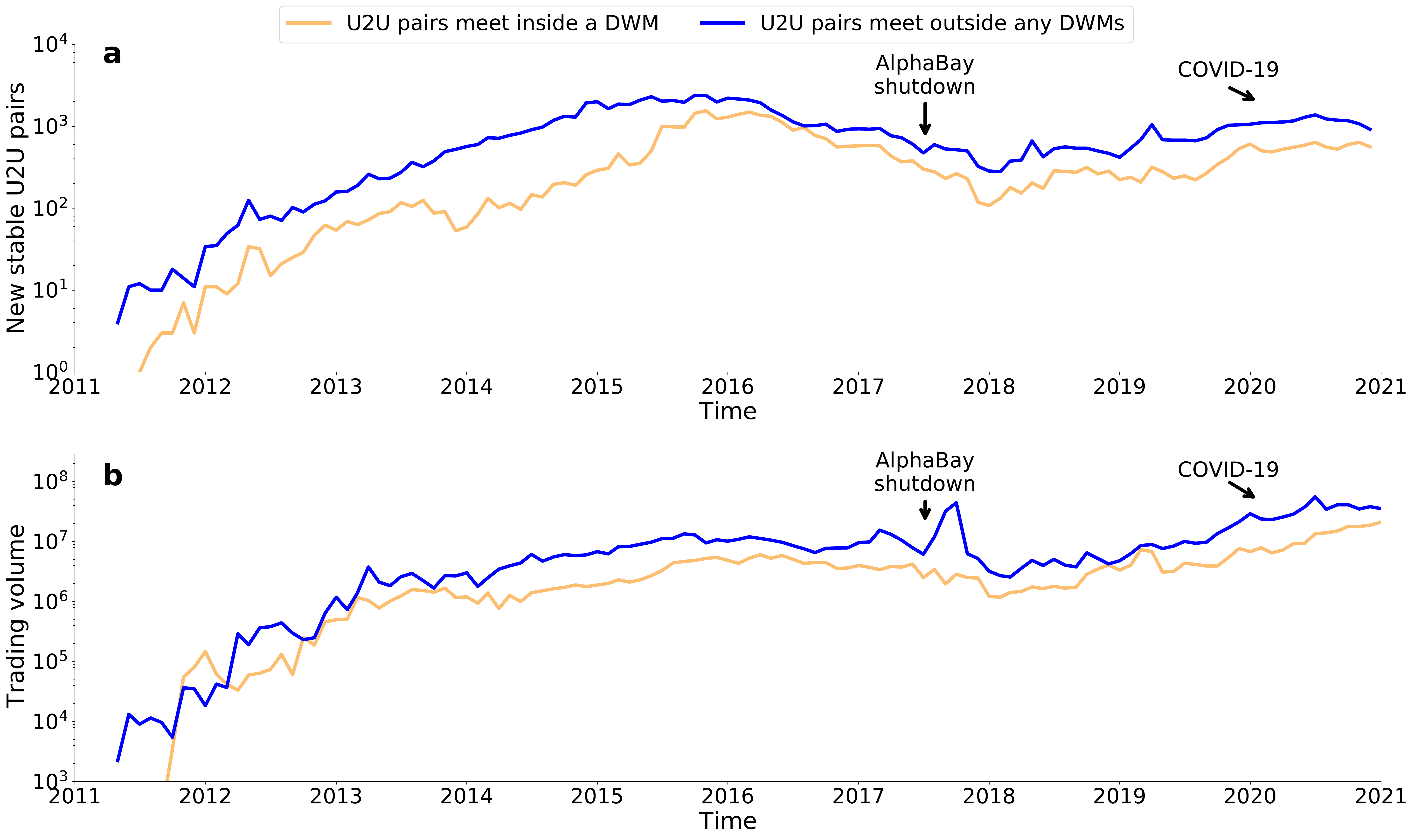}
  \caption{\textbf{Temporal evolution of stable pairs.}
  (a) Monthly number of new stable U2U pairs created. (b) Monthly trading volume of stable U2U pairs.}
  \label{Temporal_analysis}
\end{figure}

\begin{figure}[H]
  \centering
  \includegraphics[width=16cm]{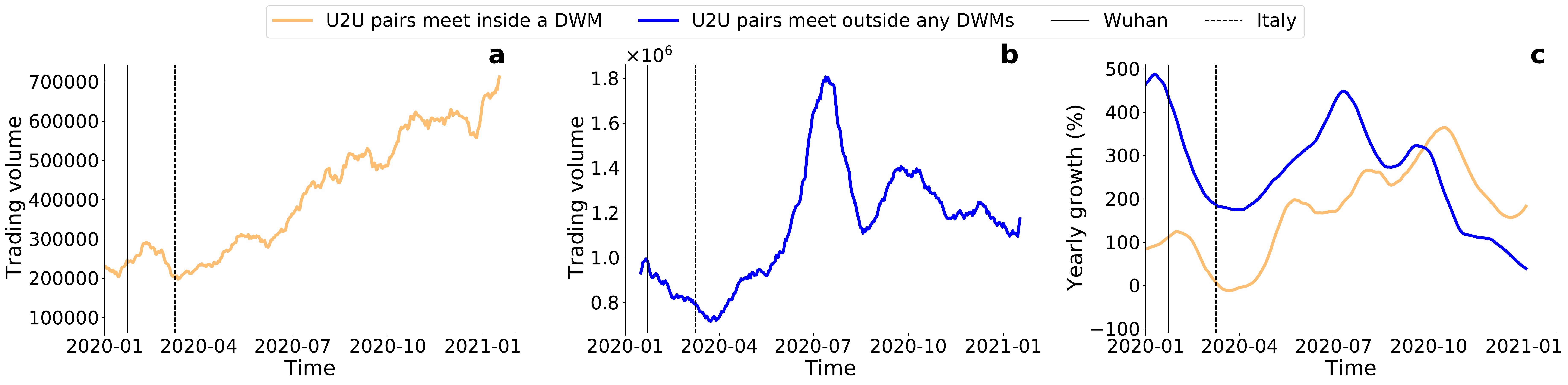}
   \caption{\textbf{Trading volume stable U2U pairs during COVID-19.} 28-days moving average of trading volume between users who met inside a DWM (a) and outside any DWMs (b). (c) Yearly growth relative to the same day of 2019. Vertical lines represent the dates of Wuhan (Jan 23, 2020) and Italy (March 3, 2020) lockdowns.}
  \label{Trading_volume_COVID}
\end{figure}

\begin{figure}[H]
  \centering
  \includegraphics[width=16cm]{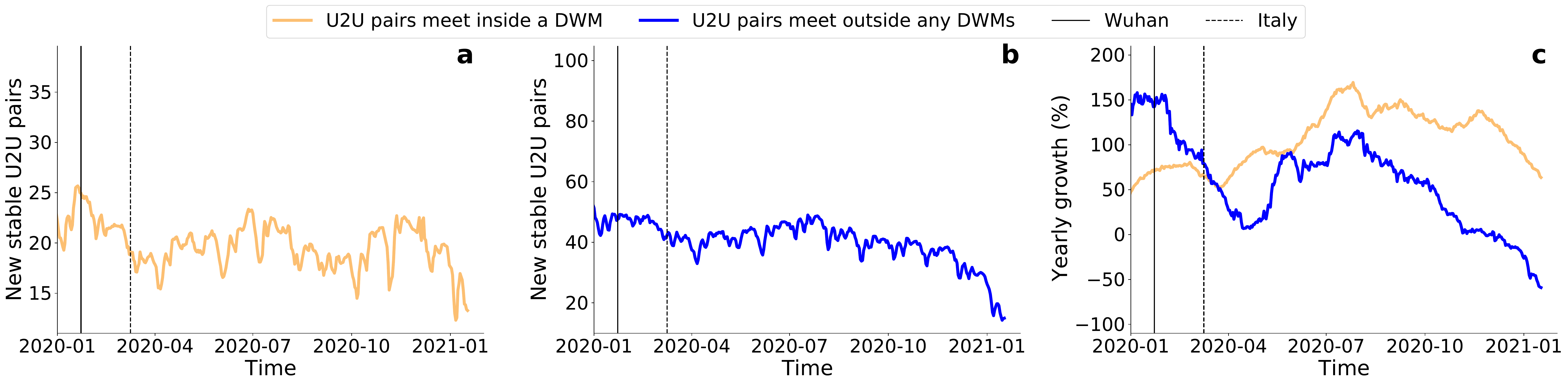}
   \caption{\textbf{Formation of new stable U2U pairs during COVID-19.} 28-days moving average of new stable U2U pairs started between U2U pairs who met inside a DWM (a) and outside any DWMs (b). (c) Yearly growth relative to the same day of 2019. Vertical lines represent the dates of Wuhan and Italy lockdowns.}
  \label{new_U2U_pairs_COVID19}
\end{figure}

We have shown that the U2U network is resilient to short-lasting external shocks, namely the closure of a marketplace, and it does not need the centralised structure of DWMs to survive. What about long-lasting systemic stress? To answer this question, we consider the impact that the COVID-19 pandemic has had on the evolution of stable U2U pairs. Previous studies reported that COVID-19 had a strong impact on DWMs, with reported delays and damage to the shipping infrastructure due to border closures~\cite{bergeron2020preliminary, Chainalysis_DWMs_growth_2020}. We start by investigating the number of new stable U2U pairs and their trading volume. Users in stable pairs meeting both inside and outside DWMs have been growing over the last two years. In 2020, a total of 6,778 pairs of users in stable pairs met inside a DWM, corresponding to the 192\% of 2019 and to the 255\% of 2018, see Figure~\ref{Temporal_analysis}(a). Pairs of users in stable pairs meeting inside a DWM traded for a total of \$145 million in 2020, which corresponds to the 252\% of 2019, and the 593\% of 2018, see Figure~\ref{Temporal_analysis}(b). 
We see similar trends for stable U2U pairs meeting outside any DWMs. The impact of the COVID-19 pandemic has, however, had different phases, determined by the number and level of measures introduced around the world. For users in stable pairs who met both inside or outside DWMs, we find that during the first lockdowns in 2020 trading volume fell with respect to January of the same year, suggesting that they were negatively impacted by COVID-19 restrictions. After that, trading volume sharply increased over all 2020, see Figure~\ref{Trading_volume_COVID}. The number of stable U2U pairs created each day was, however, steady over time during 2020, even though more U2U pairs were created compared to the same period of 2019, see Figure~\ref{new_U2U_pairs_COVID19}. Overall, stable U2U pairs have shown resilience to the systemic stress caused by COVID-19, suggesting, once again, that these trading relationships are fundamentally independent from the underlying DWMs.

\section*{Discussion and Conclusion}
\label{Discussions_conclusion}

In this paper, we revealed the prevalence and structure of a large network of direct transactions between users who trade on the same DWM. We showed that some of the links of this user-to-user (U2U) network are ephemeral while other persist in time. We highlighted that a significant fraction of stable U2U pairs formed as their members were trading with the same DWM, suggesting that DWMs may play a role in promoting the formation of stable U2U pairs. We showed that the relationships between users forming stable pairs persist even after the DWM shuts down and are not significantly affected by COVID-19, suggesting overall resilience of stable pairs to external shocks.

Our study has several limitations. In particular, our dataset does not include any attributes related to either users or their Bitcoin transactions, such as, whether the transaction represents an actual purchase or not. Moreover, we do not have information about which users trade with other users on the same DWM. Finally, our coverage of DWMs, albeit extensive, may lack information on other DWMs where users could have met. 

Our work has several policy implications. Our findings suggest that DWMs are much more than mere marketplaces~\cite{gupta2021dark}. DWMs are also communication platforms, where users can meet and chat with other users either directly -- using Whatsapp, phone, or email -- or through specialised forums. These direct interactions may favour the emergence of decentralised trade networks that bypass the intermediary role of the marketplace, similar to what is currently happening on Facebook, Telegram, and Reddit~\cite{oksanen2020illicit, bakken2019sellers,DarknetLive_telegram, sung2021prevalence, childs2021beyond, kwon2021dark}, where users post products, negotiate item prices, and then trade directly without an intermediary. We estimate that the trading volume of U2U pairs meeting on DWMs is increasing, reaching a peak in 2020 (during the COVID-19 pandemic). Indeed, our results support recent recommendations of paying attention to single sellers rather than entire DWMs~\cite{Cryptomarkets2020}. Law enforcement agencies, however, have only recently started targeting single sellers. The first operation took place in 2018 and successfully led to the arrest of 35 sellers~\cite{FirstLargeVendorArrest2018}, while the largest operation to date occurred in 2020 and led to 179 arrests in six different countries~\cite{Europol2020}. Our study indicates that a much higher number of highly active DWM users, on the order of tens of thousands, is involved in transactions with other DWM users.

Overall, our study provides a first step towards the understanding of how users of DWMs collectively behave outside organised marketplace. We believe that the results might suggest to researchers, practitioners, and law enforcement agencies that a shift in the attention from the evolution of DWMs to the behaviour of their users might facilitate the design of more appropriate strategies to counteract online trading of illicit goods.

\section*{Competing interests}
The authors declare that they have no competing interests.

\section*{Author's contributions}
M.N., A.Br., A.E., P.G., A.T., and A.Ba. designed the research; A.E. and P.G., acquired, prepared, and cleansed the data. M.N. and A.Br. performed the measurements. M.N., A.Br., A.E., P.G., A.T., and A.Ba. analysed the data. M.N., A.Br., P.G., A.T., and 
A.Ba. wrote the manuscript. M.N., A.Br., A.E., P.G., A.T., and A.Ba. discussed the results and commented on the manuscript.

\section*{Acknowledgements}
M.N., A.Br., A.T., and A.Ba. were supported by ESRC as part of UK Research and Innovation’s rapid response to COVID-19, through grant ES/V00400X/1.

\section*{Data availability}
All data needed to evaluate the conclusions in the paper are present in the paper. Additional data related to this paper may be requested from the authors.

\section*{Data and methods}
\label{Data_methods}

Additional considerations on our data and methods are available in Section~\ref{Data_methods_SI}.

\paragraph{Data preprocessing.} 
We consider only a subset of the transactions in our dataset. Namely, the ones made by the 40 entities representing the 40 DWMs under consideration, which directly interact with more than 16 million other entities, who are the users of these DWMs. Users interacting with other users form U2U pairs and we include them in our dataset. We instead discard single Bitcoin transactions below \$0.01 or above \$100,000, which are unlikely to show real purchases and minimise false positives. They may be attributed to a residual amount of Bitcoins in an address or transactions between two business partners where no good is actually given in return, respectively. The analysed dataset includes about 31 million transactions among more than 16 million users. Finally, we note that the same user can interact in multiple DWMs~\cite{elbahrawy2020collective, hiramoto2020measuring}. By definition, users that interact among themselves form U2U transactions. If the pair of users interact with multiple DWMs these U2U transactions are included in all relative DWMs and counted multiple times. Therefore, the simple sum of all U2U transactions of each DWM is more than the sum of all unique U2U transactions. We count a total of 11 million transactions around all DWMs, that goes down to 9.9 million when multiple counting is avoided. Similarly, the simple sum of the single trading volumes surrounding all DWMs amounts to \$33 billion, while the overall trading volume in all unique U2U pairs is \$30 billion. Among the 40 large DWMs under consideration, 17 participate in at least one transaction in either 2020 or 2021, while the remaining 23 closed before 2020. Notably, our dataset includes Silk Road (the first modern DWM)~\cite{christin2013traveling}, Alphabay (once the leading DWM)~\cite{van2016drugs}, and Hydra (currently the largest DWM in Russia)~\cite{elbahrawy2020collective}. Other general statistics about our dataset can be found in the Section~\ref{general_statistics_label}.

\begin{figure}[H]
  \centering
\includegraphics[width=0.7\textwidth]{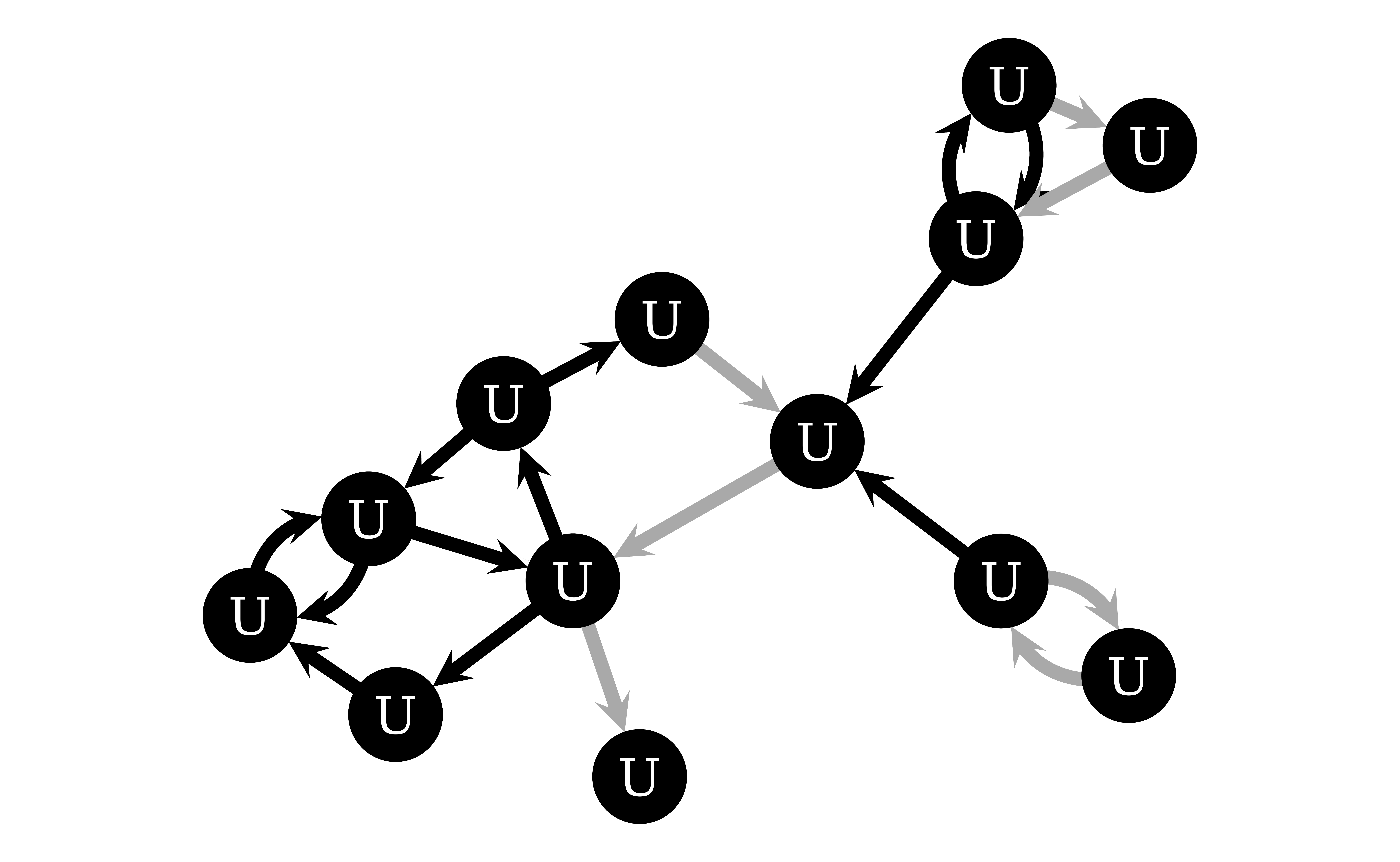}
  \caption{\textbf{U2U network.} The U2U network is formed by the entire set of interacting users (black and gray arrows with their respective users). Using the evolving activity-driven model~\cite{nadini2020detecting}, U2U pairs are divided in either stable (black arrows and respective users) or non-stable (gray arrows and respective users).}
  \label{Backbone_U2U_network}
\end{figure}

\paragraph{Detection of the U2U network.} The detection of stable U2U pairs in the full network is done by using the evolving activity-driven model~\cite{nadini2020detecting}, which introduced a statistically-principled methodology to detect the network backbone against what expected from a proper null model. If a U2U pair occurs significantly more than what expected from the null model, it is labeled as stable, otherwise as non-stable. The evolving activity-driven model is an appropriate methodology for large temporal networks~\cite{nadini2020reconstructing} and it is implemented in the Python 3 pip library TemporalBackbone~\cite{TemporalBackbone_pip2021}, where default parameter values have been used. As input parameter, we considered the full network, comprehending transactions from/to DWMs and U2U transactions between users (see Section~\ref{Detecting_trading_partners}).

\paragraph{Users who met inside a DWM.} We determine whether U2U pairs meet while active on a DWM by looking at the time occurrence of their first U2U transaction. This transaction can occur at three different moment in time. (i) At $t=t_1$, before both users interact with the same DWM (occurring at $t=t_2>t_1$ and $t=t_3>t_1$, respectively), as shown on the left hand side of Table~\ref{Motifs_Triads_table11}. (ii) At $t=t_2$, when only one user has interacted with a specific DWM and the other user will do so at a later time, as in the middle column of Table~\ref{Motifs_Triads_table11}. (iii) At $t=t_3$, when both users have interacted with the same DWM, as in the right column of Table~\ref{Motifs_Triads_table11}. We classify these three chain of events in two groups. One group includes all pairs that meet outside any DWMs, which includes case (i) and case (ii), and the other group users that meet inside a DWM, described by case (iii). This last case constitute a conservative proxy for users that meet who met inside aa DWM. The proxy admits the possibility of false positives, since it consider users who met inside a the same DWM without having interacted on it, as well as false negatives, since it does not take into account users who met inside a DWM without having ever interacted on it. The latter is arguably more significant, since it is possible that only one of the two users (the seller) has actually engaged in transactions with the DWM, while the other user, after seeing the seller’s profile on a DWM, has established a direct contact, through Whatsapp, email, or phone. 

\paragraph{Nomenclature of all groups considered.} We provide the definition of all considered groups in Table~\ref{Nomenclature}. 

\begin{table}[H]
\centering
\begin{adjustbox}{max width=\textwidth}
\begin{tabular}{l|l|c}
Group & Description & Number of users \\
\hline
\thead{1. Users who do not form \\ stable U2U pairs} & \thead{Users that form neither stable \\ U2U pair nor U2U pairs} & 15,875,877 \\
\thead{2. Users who form \\ stable U2U pairs} & \thead{ Users who form at least one stable U2U pair \\ as detected by our chosen metholodogy~\cite{nadini2020detecting} } &  106,648 \\
\thead{2a. Users who met \\ outside DWMs} & \thead{Users that form stable pairs and met at least \\ one other user following the chain \\ of events in Table~\ref{Motifs_Triads_table11} (first two columns)} & 88,828  \\
\thead{2b. Users who met \\ inside a DWM} & \thead{Users that form stable pairs and met at least \\ one other user following the chain \\ of events in Table~\ref{Motifs_Triads_table11} (last column)} &  37,129 \\
\end{tabular}
\end{adjustbox}
\caption{\textbf{Nomenclature.} Definitions of all groups the users are divided to based on their behaviour. Number of users in each group is given in the last column.}
\label{Nomenclature}
\end{table}

\bibliographystyle{unsrt}

\clearpage

\begin{center}
\LARGE{\textbf{Supplementary Information}}
\end{center}

\section{Additional data and methods}
\label{Data_methods_SI}

\paragraph{Identification of real identities performing Bitcoin transactions.} 
The trading volume of DWMs has been steadily increasing and exceeded \$1.5 billion for the first time in 2020~\cite{Chainalysis_crypto_crime_report_2021}. The vast majority of such trading has occurred in Bitcoin, which is the most popular cryptocurrency to date. Its worldwide adoption has further increased in 2021, jumping over 880\% with respect to 2020~\cite{Chainalysis_adoption}. Bitcoin allows users to use pseudonym (public address) instead of their real identities. Users can create a new pseudonym at each transaction, requiring only a computer and an internet connection. However, various heuristics exist to cluster addresses together to recover the real identity behind pseudonyms~\cite{khalilov2018survey}. In our dataset, this process is done by Chainalysis Inc. (see Section~\ref{dark_web_anonymity}). In the dataset, real entities represent DWMs, users of DWMs, or other entities interacting with these users. Transactions to and from Bitcoin trading exchanges are removed, because our primary interest entails the study of direct interactions between DWMs and single users. The dataset comprises of 40 DWMs, for a total of 149 transactions among 57 million real entities. Each Bitcoin transaction has an associated timestamp $t$, indicating the time at which the transaction occurred. The dataset is sparse, with 54.6\% of all entities performing a transaction only. The conversion from Bitcoin to dollars is done using the price of Bitcoin at the time of the transaction.

\paragraph{Evaluation of coefficients of the trend line in Figure~\ref{Importance_U2U_transactions}(a).} The coefficients $a=1.06$ and $b = 0.70$ of the trend line $y = x^a 10^{-b}$ in Figure~\ref{Importance_U2U_transactions}(a) are in good agreement with the empirical data, $R^2 = 0.969$, and evaluated as follows. First, the equation is transformed to $Y = a X - b$, where $Y = \log_{10} y$ and $X = \log_{10} x $. The linear equation fitted against real data and coefficients $a$ and $b$ computed by minimizing the sum of squares. 

\paragraph{Statistical analysis.} We compare the median of two paired distributions using the two-sided Wilcoxon test~\cite{wilcoxon1992individual}. It is a non-parametric statistical test and verifies the null hypothesis that two paired samples come from distributions with the same median. If distributions are not paired, we use the Mann-Whitney-U test to assess statistical differences of the medians of two distributions~\cite{mann1947test}. We compare two distributions using the Kolmogorov-Smirnov test~\cite{massey1951kolmogorov} on two samples. It tests the null hypothesis that 2 independent samples are drawn from the same continuous distribution. We evaluate the correlation between two sets of values using the Spearman rank-order correlation coefficient~\cite{spearman1961proof}. It is a correlation coefficient that does not assume normally distributed values and varies between -1 and 1: with -1 implying a negative correlation, 0 no correlation, and 1 a positive correlation.

\section{Dark web marketplaces and identification of real identities performing Bitcoin transactions.}
\label{dark_web_anonymity}

\subsection{Dark web marketplaces}
DWMs are in many ways similar to other online marketplaces. They have strict policies that every user must follow. For instance, in some DWMs are banned categories of products, like human trafficking, contract killing, weapons, or COVID-19 fake vaccines~\cite{DarkNetLive_ban, monopoly_ban}. Registration is required for all sellers, and sometimes also for buyers. Certified sellers can advertise their products. They have a reputation, which is based on buyers' reviews~\cite{wehinger2011dark, soska2015measuring}. They are also responsible for delivering the products, sometimes with a tracking number attached, and may offer refunds or reshipment. Buyers are free to look at the listings and sometimes can ask questions directly to the relative seller~\cite{COVID19Drugs, bracci2020covid}. Payments are often protected by escrow services. These are third-party services, which guarantee that buyers can safely have their money refunded. Users' on DWMs constitute an active community. Numerous are websites and forums where users can share their experience and get advice on the most trustworthy DWMs and sellers, such as Dread~\cite{Dread}, Raptor.life~\cite{Raptor.life}, DarkNetLive~\cite{DarkNetLive}, and DarkFail~\cite{DarkFail}.

DWMs have some unique features as well. They sell several kinds of illicit products, like drugs, fake IDs, and medicines~\cite{barratt2014use, martin2014lost, aldridge2014not}. They are not accessible by standard web search-engines, but operate online in an encrypted part of the Internet~\cite{martin2014drugs}. Potential buyers can easily access to DWMs using specialized browsers, like The Onion Router (Tor)~\cite{dingledine2004tor}, and anonymously trade illicit goods using cryptocurrencies, like Bitcoin~\cite{nakamoto2008Bitcoin}. Bitcoin is currently the most popular cryptocurrency on DWMs~\cite{lee2019cybercriminal, foley2019sex, moser2018empirical} and its adoption is growing in the regular economy as well. Its infrastructure seems to ensure complete anonymity to its users. If a proper technique is adopted, however, there are chances to link the Bitcoin blockchain (that is, the entire Bitcoin transaction history) with the user's real identity~\cite{khalilov2018survey}. When the Bitcoin blockchain is successfully linked to a real identity, the records of past, present, and future Bitcoin transactions is traceable, easily accessible, and can be used by companies, law enforcement agencies, and researchers.

\subsection{Identification of real identities performing Bitcoin transactions}
The raw, anonymized Bitcoin blockchain can be publicly accessed through Bitcoin core~\cite{Bitcoin_core} or third-party APIs such as Blockchain.com~\cite{blockchain}. It contains information about origin and destination addresses, as well as time and amount of the transactions. In order to contrast traceability of the real identity, an user is likely to use multiple addresses. A new address is often generated in each transaction. Grouping the addresses in clusters reduces the complexity of the Bitcoin blockchain and challenge users' anonymity~\cite{ron2013quantitative}. Given that millions of Bitcoin addresses are currently active and many others are continuously being generated, a clustering approach primarily based on manual annotation is not feasible. Various heuristics, instead, have been proposed\cite{ron2013quantitative, androulaki2013evaluating,   tasca2018evolution, harrigan2016unreasonable}. They were successful in grouping Bitcoin addresses and associate them to cluster of real entities. For instance, in~\cite{ron2013quantitative}, the authors were able to find a connection between a set of large transactions and a single one, which was dated in November 2010. In~\cite{androulaki2013evaluating}, the authors applied to a daily university setting the privacy protocol recommended in Bitcoin transactions, finding that almost 40\% of the real identities would be recovered. Another work showed the presence of ``super clusters'' of entities, which marked macro-variations in the evolution of the Bitcoin economy~\cite{tasca2018evolution}. The primary reasons behind the effectiveness of heuristic clustering are: ``address reuse, avoidable merging, super-clusters with high centrality, and the incremental growth of address clusters''~\cite{harrigan2016unreasonable}.

\begin{figure}[H]
  \centering
 \includegraphics[width=10cm]{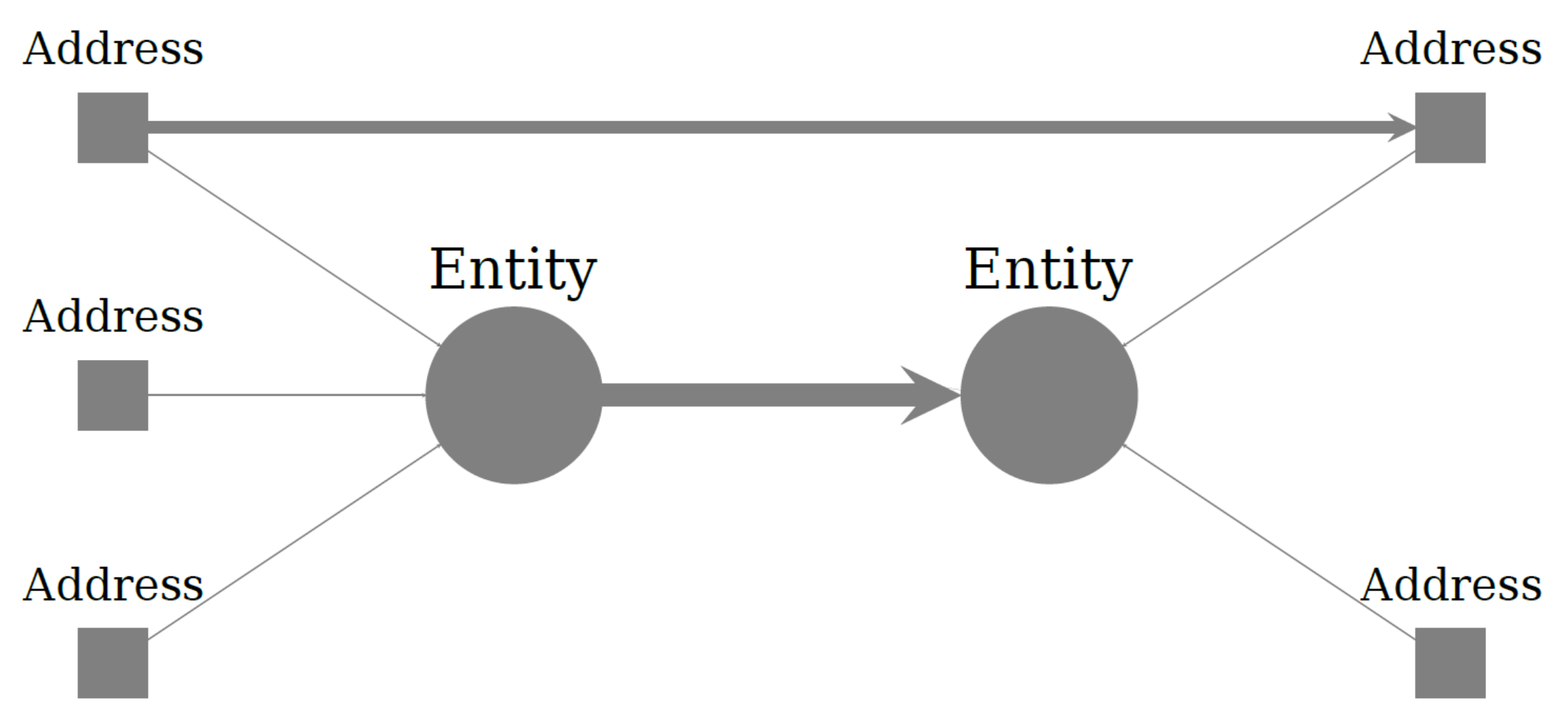}
  \caption{\textbf{Identification of real entities in the Blockchain.} End goal of Bitcoin transactions clustering techniques: mapping a series of Bitcoin addresses to real entities. In this example, an address sends Bitcoins to another address. Thanks to the identification process, the two addresses are associated with two real entities. The Bitcoin transaction between the two entities becomes traceable and transparent.}
  \label{From_addresses_to_entities}
\end{figure}

The end goal of clustering Bitcoin addresses is to map them to single, real entities, as shown in Figure~\ref{From_addresses_to_entities}. To achieve this goal, however, heuristic clustering techniques should be improved. Manual annotation has shown a valuable potential~\cite{meiklejohn2013fistful}. It consists on gathering publicly available Bitcoin addresses, like the Wikimedia Foundation one~\cite{Coindesk}, and engage through direct interaction with unknown Bitcoin addresses. If some real entities are known, it is easier to associate the remaining Bitcoin addresses to other real identities. In the last few years, companies specialising in Bitcoin analytics have started to leverage previous methodologies~\cite{ron2013quantitative, androulaki2013evaluating, tasca2018evolution, harrigan2016unreasonable, meiklejohn2013fistful} to unveil real entities. The leading company in analysing Bitcoin transactions on DWMs is Chainalysis Inc.~\cite{Chainalysis}, which has also aided several federal investigations. For instance, it supported the United States Internal Revenue Service (IRS) in tracking Bitcoin transactions~\cite{chung2019cracking} and the FBI in the Twitter hack~\cite{Chainalysis_twitter_hack}.  Chainalsysis clusters Bitcoin transactions in groups by combining previous methodologies~\cite{ron2013quantitative, androulaki2013evaluating, tasca2018evolution, harrigan2016unreasonable} and real entities are unveiled with an approach similar to~\cite{meiklejohn2013fistful} (see Section~\ref{dark_web_anonymity} for more details on DWMs and this clustering technique). In the dataset, real entities represent DWMs, users of DWMs, or other real entities interacting with these users. Chainalysis aims at minimizing the false positives, who may lead to wrongly associate a real entity with illicit activities. If a Bitcoin address cannot be uniquely ascribed to a real entity, it is included in our dataset as an independent and unnamed entity. Only a fraction of the entities in our dataset thus represent named and real entities, which identity is known. Given that there are millions of entities in our dataset, it is impossible to identify all the corresponding real identities. After the identification process is completed, to each real entity is associated a string of numbers and the dataset re-anonymized. Transactions to and from Bitcoin trading exchanges are also removed, because our primary interest entails the study of direct interactions between real entities.  

\clearpage

\section{General statistics of the 40 DWMs under consideration}
\label{general_statistics_label}

\begin{table}[H]
\begin{adjustbox}{max width=\textwidth}
\begin{tabular}{c|cc|cc}
\multirow{2}{*}{Name}   & \multicolumn{2}{c|}{Transactions with a DWM}  & \multicolumn{2}{c}{U2U transactions}  \\
 & \thead{Users \\ (sent; received; total)} & \thead{Trading volume\\in millions\\ (sent; received; total)} & \thead{Users \\ (sent; received; total)} & \thead{Trading volume\\in millions\\ (total)} \\
\hline
\hline
\thead{Abraxas}	&	\thead{(95,642; 21,500; 111,003)}	&	\thead{(21.85; 27.23; 49.09)}	&	\thead{(28,588; 25,546; 44,151)}	&	\thead{61.92}	 \\ 
\thead{Agora}	&	\thead{(462,106; 119,221; 537,983)}	&	\thead{(141.3; 132.8; 274.1)}	&	\thead{(168,248; 151,699; 252,984)}	&	\thead{558.0}	 \\ 
\thead{AlphaBay}	&	\thead{(1,658,059; 334,154; 1,898,850)}	&	\thead{(537.1; 568.6; 1,106)}	&	\thead{(524,783; 422,881; 776,183)}	&	\thead{1581}	 \\ 
\thead{Apollon}	&	\thead{(68,373; 13,954; 79,307)}	&	\thead{(12.90; 16.59; 29.50)}	&	\thead{(19,468; 17,290; 29,900)}	&	\thead{49.38}	 \\ 
\thead{Basetools}	&	\thead{(119,114; 347; 119,461)}	&	\thead{(4.712; 6.727; 11.44)}	&	\thead{(32,191; 34,169; 50,939)}	&	\thead{63.23}	 \\ 
\thead{Benumb Shop}	&	\thead{(27,229; 343; 27,556)}	&	\thead{(3.929; 5.027; 8.956)}	&	\thead{(5,499; 5,654; 8,985)}	&	\thead{21.73}	 \\ 
\thead{BitBazaar}	&	\thead{(20,805; 150; 20,931)}	&	\thead{(2.681; 4.425; 7.106)}	&	\thead{(6,939; 6,569; 10,126)}	&	\thead{14.13}	 \\ 
\thead{Black Bank}	&	\thead{(52,783; 15,147; 64,131)}	&	\thead{(11.41; 11.78; 23.19)}	&	\thead{(15,843; 13,486; 24,291)}	&	\thead{31.11}	 \\ 
\thead{Blue Sky}	&	\thead{(16,002; 10,140; 22,616)}	&	\thead{(3.225; 3.786; 7.011)}	&	\thead{(9,763; 6,149; 12,108)}	&	\thead{10.86}	 \\ 
\thead{Buybest}	&	\thead{(334,741; 3,004; 337,556)}	&	\thead{(24.45; 7.490; 31.94)}	&	\thead{(57,001; 59,131; 99,390)}	&	\thead{132.4}	 \\ 
\thead{Bypass Shop}	&	\thead{(861,716; 8,118; 869,593)}	&	\thead{(65.66; 54.36; 120.0)}	&	\thead{(176,905; 174,151; 288,745)}	&	\thead{804.3}	 \\ 
\thead{DarkMarket}	&	\thead{(176,141; 13,554; 183,010)}	&	\thead{(27.25; 36.62; 63.87)}	&	\thead{(72,923; 67,621; 105,416)}	&	\thead{166.7}	 \\ 
\thead{Dream}	&	\thead{(466,511; 45,399; 507,837)}	&	\thead{(57.64; 72.95; 130.6)}	&	\thead{(109,871; 70,706; 154,873)}	&	\thead{287.3}	 \\ 
\thead{Empire}	&	\thead{(405,202; 9,690; 413,858)}	&	\thead{(64.38; 56.84; 121.2)}	&	\thead{(63,431; 56,415; 103,886)}	&	\thead{287.4}	 \\ 
\thead{Evolution}	&	\thead{(216,604; 34,512; 240,713)}	&	\thead{(47.58; 50.15; 97.73)}	&	\thead{(77,331; 72,711; 115,496)}	&	\thead{236.7}	 \\ 
\thead{FEshop}	&	\thead{(1,134,456; 5,858; 1,140,275)}	&	\thead{(64.67; 48.83; 113.5)}	&	\thead{(244,318; 261,489; 420,040)}	&	\thead{834.4}	 \\ 
\thead{Flugsvamp 2.0}	&	\thead{(104,385; 21,201; 119,893)}	&	\thead{(23.01; 38.20; 61.22)}	&	\thead{(29,215; 23,079; 41,047)}	&	\thead{144.2}	 \\ 
\thead{Flugsvamp 3.0}	&	\thead{(217,083; 20,773; 234,563)}	&	\thead{(39.34; 52.78; 92.12)}	&	\thead{(52,075; 49,881; 81,527)}	&	\thead{473.8}	 \\ 
\thead{FuLLzShOp}	&	\thead{(21,716; 9; 21,726)}	&	\thead{(3.937; 4.510; 8.447)}	&	\thead{(4,147; 4,496; 7,209)}	&	\thead{10.07}	 \\ 
\thead{Hansa}	&	\thead{(330,565; 73,202; 358,120)}	&	\thead{(60.64; 55.91; 116.6)}	&	\thead{(153,567; 127,514; 209,717)}	&	\thead{76.60}	 \\ 
\thead{Hydra}	&	\thead{(4,031,013; 666,075; 4,584,339)}	&	\thead{(1,868; 1,810; 3,678)}	&	\thead{(2,447,548; 2,099,320; 3,124,366)}	&	\thead{20,840}	 \\ 
\thead{Joker's Stash}	&	\thead{(806,089; 1,090; 807,140)}	&	\thead{(153.0; 49.95; 203.0)}	&	\thead{(154,872; 156,689; 260,832)}	&	\thead{926.2}	 \\ 
\thead{LuxSocks.ru}	&	\thead{(326,159; 186; 326,340)}	&	\thead{(8.123; 5.573; 13.70)}	&	\thead{(59,638; 66,011; 97,705)}	&	\thead{175.8}	 \\ 
\thead{Matanga}	&	\thead{(57,354; 633; 57,963)}	&	\thead{(5.882; 7.775; 13.66)}	&	\thead{(10,637; 10,328; 17,632)}	&	\thead{96.35}	 \\ 
\thead{Middle Earth}	&	\thead{(38,017; 9,206; 45,629)}	&	\thead{(8.361; 9.151; 17.51)}	&	\thead{(8,091; 7,603; 12,990)}	&	\thead{18.68}	 \\ 
\thead{MrGreen.ws}	&	\thead{(44,918; 176; 45,094)}	&	\thead{(8.244; 6.176; 14.42)}	&	\thead{(6,298; 5,912; 10,501)}	&	\thead{14.44}	 \\ 
\thead{Nightmare}	&	\thead{(37,844; 3,524; 40,894)}	&	\thead{(5.697; 7.371; 13.07)}	&	\thead{(8,830; 6,277; 12,905)}	&	\thead{25.83}	 \\ 
\end{tabular}
\end{adjustbox}
\caption{\textbf{General statistics of DWMs, part 1.} Some DWMs are presented here, the others are available in Table~\ref{General_statistics_2}. The terms ``sent'' and ``received'' always refer to transactions made by users. The trading volume indicates millions of dollars. The amount of dollars sent and received by users through U2U transactions is equivalent to the total.}
\label{General_statistics_1}
\end{table}

\begin{table}[H]
\begin{adjustbox}{max width=\textwidth}
\begin{tabular}{c|cc|cc}
\multirow{2}{*}{Name}   & \multicolumn{2}{c|}{Interactions with DWM}  & \multicolumn{2}{c}{U2U interactions}  \\
 & \thead{Users \\ (sent; received; total)} & \thead{Trading volume\\in millions\\ (sent; received; total)} & \thead{Users \\ (sent; received; total)} & \thead{Trading volume\\in millions\\ (total)} \\
\hline
\hline 
\thead{Nucleus}	&	\thead{(205,043; 53,571; 247,884)}	&	\thead{(56.59; 61.68; 118.3)}	&	\thead{(62,577; 52,829; 93,279)}	&	\thead{156.9}	 \\ 
\thead{Pandora}	&	\thead{(35,667; 8,723; 41,718)}	&	\thead{(8.401; 8.561; 16.96)}	&	\thead{(11,119; 8,964; 15,944)}	&	\thead{26.37}	 \\ 
\thead{Russian Anonymous}	&	\thead{(740,625; 36,161; 769,228)}	&	\thead{(80.24; 95.94; 176.2)}	&	\thead{(363,773; 331,811; 493,766)}	&	\thead{1866}	 \\ 
\thead{San-Wells}	&	\thead{(51,795; 2,858; 54,633)}	&	\thead{(6.335; 5.755; 12.09)}	&	\thead{(8,227; 7,841; 13,679)}	&	\thead{15.36}	 \\ 
\thead{Sheep}	&	\thead{(38,068; 7,634; 42,673)}	&	\thead{(10.81; 11.47; 22.29)}	&	\thead{(12,007; 10,288; 182,90)}	&	\thead{49.87}	 \\ 
\thead{Silk Road}	&	\thead{(382,534; 72,344; 429,284)}	&	\thead{(130.2; 149.7; 279.9)}	&	\thead{(163,376; 157,113; 243,441)}	&	\thead{671.4}	 \\ 
\thead{Silk Road 2}	&	\thead{(222,666; 47,528; 254,830)}	&	\thead{(66.83; 71.92; 138.7)}	&	\thead{(73,116; 66,019; 111,387)}	&	\thead{259.8}	 \\ 
\thead{Silk Road 3.1}	&	\thead{(59,894; 15,413; 70,078)}	&	\thead{(9.054; 13.49; 22.54)}	&	\thead{(22,160; 18,570; 32,491)}	&	\thead{21.80}	 \\ 
\thead{TradeRoute}	&	\thead{(103,517; 14,080; 112,634)}	&	\thead{(16.97; 17.04; 34.01)}	&	\thead{(27,901; 22,287; 41,869)}	&	\thead{67.72}	 \\ 
\thead{Unicc}	&	\thead{(2,004,236; 559; 2,004,789)}	&	\thead{(147.8; 84.61; 232.4)}	&	\thead{(473,969; 490,794; 780,282)}	&	\thead{1,673}	 \\ 
\thead{Valhalla}	&	\thead{(82,507; 8,218; 89,214)}	&	\thead{(8.933; 9.811; 18.74)}	&	\thead{(25,755; 32,687; 45,297)}	&	\thead{51.49}	 \\ 
\thead{Wall Street}	&	\thead{(334,871; 25,352; 347,842)}	&	\thead{(48.15; 53.16; 101.3)}	&	\thead{(148,262; 127,370; 203,176)}	&	\thead{163.5}	 \\ 
\thead{xDedic}	&	\thead{(27,956; 885; 28,736)}	&	\thead{(3.552; 3.838; 7.389)}	&	\thead{(4,785; 4,767; 7,685)}	&	\thead{12.70}	 \\ 
\end{tabular}
\end{adjustbox}
\caption{\textbf{General statistics of DWMs, part 2.} Some DWMs are presented here, the others are available in Table~\ref{General_statistics_1}. The terms ``sent'' and ``received'' always refer to transactions made by users. The trading volume indicates millions of dollars. The amount of dollars sent and received by users through U2U transactions is equivalent to the total.}
\label{General_statistics_2}
\end{table}

\begin{figure}[H]
  \centering
  \includegraphics[width=16cm]{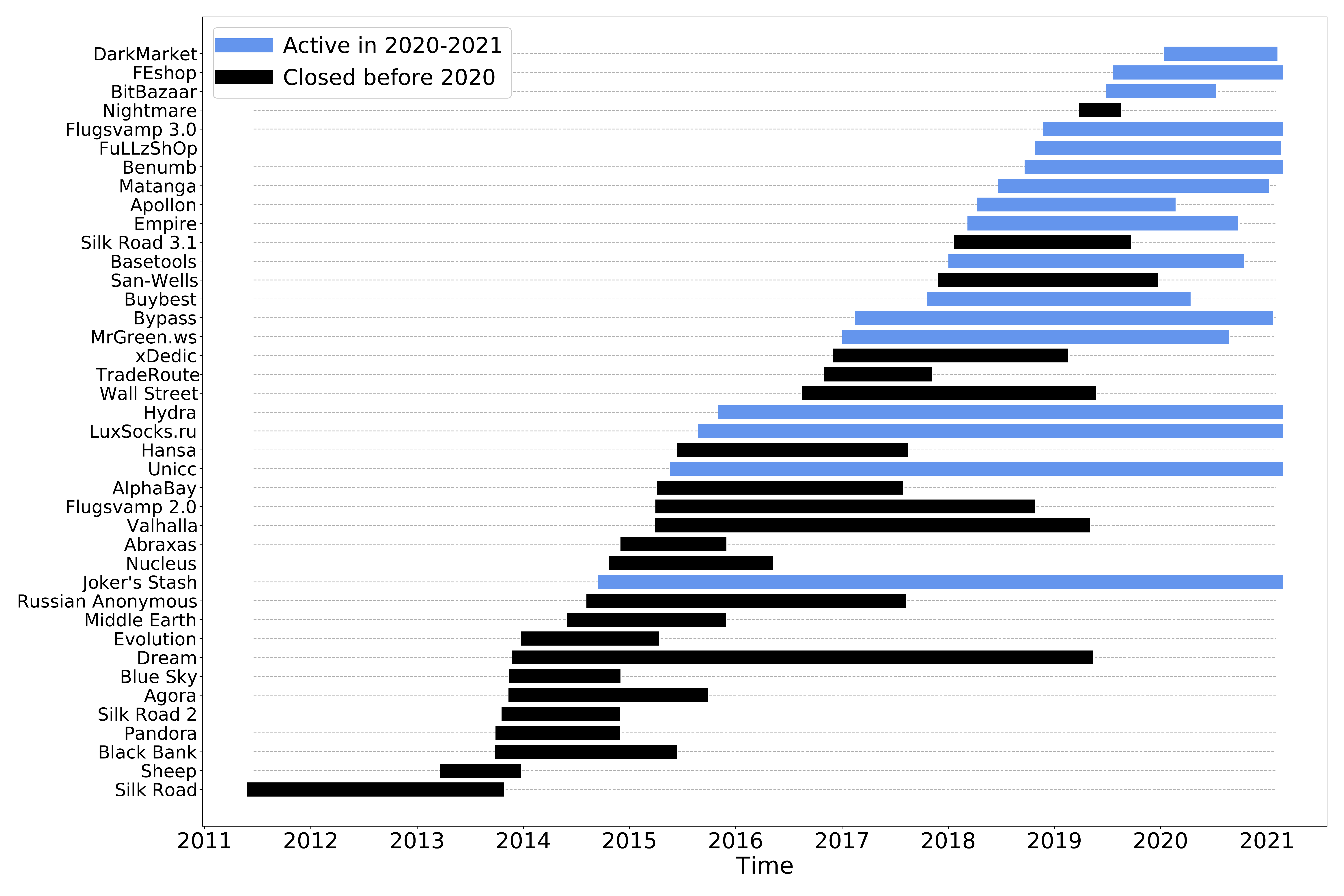}
  \caption{\textbf{Lifetime of DWMs in our dataset.} Time interval between the first and last transaction of each DWM. A total of 17 DWMs participated in at least one transactions in either 2020 or 2021, while 23 closed before 2020.}
  \label{Lifetime_DWMs_appendix}
\end{figure}

\clearpage

\section{Detection of stable pairs in temporal and directed networks}
\label{Detecting_trading_partners}

Here, we summarize the metholodogy of detecting the backbone of stable pairs in temporal and undirected networks as introduced in~\cite{nadini2020detecting}, and show how it can be easily adapted to tackle the analysis of directed temporal networks. The methodology follow three sequential steps: (i) determine the interval partition, (ii) estimate models' parameters, over successive intervals, and (iii) run a statistical filter, which removes all pairs explained by the null hypothesis and retain stable pairs. The analysed temporal network, either directed or undirected, of $N$ nodes evolves in an observation window composed of $T \gg 1$ time steps, labeled as $t=1,...,T$. At each time step $t$, entities interact among themselves and form a time-varying network of interactions, described by a binary adjacency matrix that varies in time $A(t)$.

\subsection{Temporal and undirected networks} 

\paragraph{Interval partition.} The overall observation window is divided in successive and disjoint intervals using an auxiliary method, namely, the Bayesian Block method~\cite{scargle2013studies}. It takes as input the total number of temporal pairs created in the entire network at time $t$ 
\begin{equation}
\Omega^{\text{ts}} (t) = \sum_{i,j = 1; i \leq j}^{N} A_{ij}^{\text{ts}}(t),
\label{W_t_tot}
\end{equation} 
where the superscript ``$\text{ts}$'' indicates that these variables are estimated from the time series and $A_{ij}^{\text{ts}}(t)$ is the $ij$th entry of the estimated adjacency matrix at time $t$. The Bayesian Block method returns the interval partition, which divides the overall time window $T$ into $I$ disjoint intervals indexed by $\Delta = 1, \ldots, I$, that contain a uniform total number of connections.  From the knowledge of the interval partition, the length, $\tau(\Delta)$, of the generic $\Delta$th interval is obtained with the following closure relation: $\sum_{\Delta=1}^{I} \tau(\Delta) =T$.

\paragraph{Parameter estimation.} According to the null hypothesis, pair of entities $i$ and $j$ are expected to interact proportional to the their individual activities at time $t$. That is, the probability that entities $i$ and $j$ interact at time $t$ is a binomial random variable defined as 
\begin{equation}
    p_{ij}(t) \equiv a_i(t) a_j(t),
    \label{probability_random}
\end{equation} 
where $a_i(t)$ and $a_j(t)$ are piece-wise constant activities, which represent the propensity of creating interactions at time $t$. The estimation of piece-wise constant activities is carried out analysing each of the $I$ intervals separately. The activity of entity $i$ at time $t \in \left[t_{\text{in}(\Delta)}, t_{\text{in}(\Delta)}+\tau(\Delta) -1 \right]$ is computed through the following frequency count:
    \begin{equation}
    \begin{aligned}
    a_i \left( t \right) &= \frac{s_{i}^{\mathrm{ts}} \left(\Delta \right)}{\sqrt{2 W^{\mathrm{ts}}(\Delta)\tau(\Delta)}},
    \label{Estimation_activity_EADM}
    \end{aligned}
    \end{equation}
where $s_{i}^{\mathrm{ts}} \left(\Delta \right)$ and  $W^{\mathrm{ts}}(\Delta) \gg 1$ are the total number of pairs generated by entity $i$ in the $\Delta$th and the total number of temporal pairs generated in the network in the $\Delta$th interval, respectively. These variables are computed from the adjacency matrix $A^{\text{ts}}(t)$, as, $s_{i}^{\mathrm{ts}} \left(\Delta \right) = \sum_{j = 1}^{N} \sum_{t = t_{\text{in}(\Delta)}}^{t_{\text{in}}(\Delta)+\tau(\Delta)-1} A_{ij}^{\mathrm{ts}}(t)$, and $W^{\mathrm{ts}}(\Delta)=\frac{1}{2} \sum_{i = 1}^{N} s_{i}^{\mathrm{ts}}(\Delta)$. Once the activities are estimated according with Eq.~\eqref{Estimation_activity_EADM}, the probability in Eq.~\eqref{probability_random} can be calculated.

\paragraph{Statistical filter.} The statistical filter compares expected number of connections between entity $i$ and entity $j$, $\text{E} \left[\overline{w}_{ij}\right]$, with observations from the time series, $\overline{w}_{i j}^{\mathrm{ts}} = \sum_{t=1}^T A_{ij}^{\mathrm{ts}}(t)$. The expected number of connections between entities $i$ and $j$ in the overall time window $T$ is determined by the sum of the binomial random variables given in Eq.~\eqref{probability_random}
\begin{equation}
\text{E} \left[\overline{w}_{ij}\right] = \sum_{t=1}^{T} p_{ij}(t) = \sum_{\Delta = 1}^{I} \frac{s_{i}^{\mathrm{ts}}(\Delta) s_{j}^{\mathrm{ts}}(\Delta)}{2 W^{\mathrm{ts}}(\Delta) },
\label{Expected_pairs}
\end{equation}
where we have used the estimation of activity in Eq.~\eqref{Estimation_activity_EADM} and summed over all intervals. Although the sum of non-identical binomial random variables in Eq.~\eqref{Expected_pairs} is a Poisson binomial distribution, the Poisson distribution is an appropriate approximation for long time series.
The probability that the observed weight, $\overline{w}_{i j}^{\mathrm{ts}}$, could be explained by the relative expected weight, $\text{E} \left[\overline{w}_{ij}\right]$ in Eq.~\eqref{Expected_pairs}, is computed according to the cumulative function of the Poisson distribution
\begin{equation}
\begin{aligned}
\alpha_{ij} &\equiv 1 - \sum_{x = 0}^{\overline{w}_{i j}^{\mathrm{ts}}-1} P \left(x; \text{E}\left[ \overline{w}_{ij} \right] \right),
\label{p_value_EADM_ETFM}
\end{aligned}
\end{equation}
where $P \left(x; \text{E}\left[ \overline{w}_{ij} \right] \right)$ indicates the Poisson distribution with random variable $x$ and expected value $\text{E}\left[ \overline{w}_{ij} \right]$. Equation~\eqref{p_value_EADM_ETFM} represents the p-value $\alpha_{ij}$: when the p-value is below a pre-defined threshold, the pair $ij$ is significant and included in the backbone network. The same statistical test is repeated for all pairs of entities $ij$ observed at least once in the overall temporal evolution.

\subsection{Temporal and directed networks} 

With little modifications, the above methodology can be used to filter temporal and directed networks.

\paragraph{Interval partition.} The interval partition is obtained by using the Bayesian Block method as above. The total number of temporal pairs created in the entire network at time $t$ is
\begin{equation}
\Omega^{\text{ts}} (t) = \sum_{i,j = 1}^{N} A_{ij}^{\text{ts}}(t),
\label{W_t_tot_directed}
\end{equation} 
where not pairs are directed, while in Eq.~\eqref{W_t_tot} undirected, thereby explaining the different ranges in the summations. 

\paragraph{Parameter estimation.}

In directed networks, the probability that entity $i$ contacts at random entity $j$ at time $t$ is defined as 
\begin{equation}
p_{i\rightarrow j} (t) \equiv a_i (t) b_j (t).
\label{Probability_EADAM}
\end{equation}
where $a_i(t)$ is the activity of entity $i$ at time $t$ and $b_j(t)$ the attractiveness of entity $j$ at time $t$. The activity was already defined in Eq.~\eqref{probability_random}, while the attractiveness represent the propensity of receiving connections at time $t$. If $a_i(t) = b_i(t)$ $\forall i, t$ (for all entities in the network and at all time), Eq.~\eqref{Probability_EADAM} becomes equivalent to Eq.~\eqref{probability_random}. However, care should be placed in their interpretation, whereby Eq.~\eqref{Probability_EADAM} generates a directed pair from entity $i$ to entity $j$, while Eq.~\eqref{probability_random} can only lead to an undirected pair. 

In the generic $\Delta$th interval, defining the time window $t \in \left[t_{\text{in}(\Delta)}, t_{\text{in}(\Delta)}+\tau(\Delta) -1 \right]$, piece-wise constant activities and attractivenesses are estimated directly from the time series, similarly to what done in the undirected case in Eq.~\eqref{Estimation_activity_EADM} 
\begin{equation}
\begin{aligned}
a_i(t) &= \frac{s_{\text{out}, i}^{\mathrm{ts}}(\Delta)}{\sqrt{ W^{\mathrm{ts}}(\Delta)\tau(\Delta)}} \quad b_i(t) &= \frac{s_{\text{in}, i}^{\mathrm{ts}}(\Delta)}{\sqrt{ W^{\mathrm{ts}}(\Delta)\tau(\Delta)}}, 
\label{estimation_activity_attractiveness_EADAM}
\end{aligned}
\end{equation}
where $s_{\text{out}, i}^{\mathrm{ts}}(\Delta)$, $s_{\text{in}, i}^{\mathrm{ts}}(\Delta)$, and $W^{\mathrm{ts}}(\Delta) \gg 1$, are the total incoming strength of entity $i$ in the $\Delta$th interval, outgoing strength of entity $i$ in the $\Delta$th interval, and the total number of directed, temporal pairs generated in the network in the $\Delta$th interval, respectively. These variables are computed from the adjacency matrix $A^{\text{ts}}(t)$, that is, $s_{\text{out}, i}^{\mathrm{ts}} \left(\Delta \right) = \sum_{j = 1}^{N} \sum_{t = t_{\text{in}(\Delta)}}^{t_{\text{in}}(\Delta)+\tau(\Delta)-1} A_{ij}^{\mathrm{ts}}(t)$, $s_{\text{in}, i}^{\mathrm{ts}} \left(\Delta \right) = \sum_{i = 1}^{N} \sum_{t = t_{\text{in}(\Delta)}}^{t_{\text{in}}(\Delta)+\tau(\Delta)-1} A_{ij}^{\mathrm{ts}}(t)$, and $W^{\mathrm{ts}}(\Delta)= \sum_{i = 1}^{N} s_{\text{out}, i}^{\mathrm{ts}}(\Delta)$. Once the activity and attractiveness are estimated according with Eq.~\eqref{estimation_activity_attractiveness_EADAM}, the probability in Eq.~\eqref{Probability_EADAM} can be evaluated. 

\paragraph{Statistical filter.} Similar to Eq.~\eqref{Expected_pairs}, the expected number of pairs from entity $i$ to entity $j$ is computed by summing the probability in Eq.~\eqref{Probability_EADAM} for all time instants $t$
\begin{equation}
\begin{aligned}
\text{E} \left[\overline{w}_{i \rightarrow j} \right] =& \sum_{t = 1}^{T}  p_{i\rightarrow j} (t) = \sum_{\Delta = 1}^{I} \frac{s_{\text{out}, i}^{\mathrm{ts}}(\Delta) s_{\text{in}, j}^{\mathrm{ts}}(\Delta)}{ W^{\mathrm{ts}}(\Delta) }.
\label{Expected_value_EADAM}
\end{aligned}
\end{equation}

The probability that the observed weight, $\overline{w}_{i \rightarrow j}^{\mathrm{ts}}$, is explained by the expected weight, $\text{E} \left[\overline{w}_{i \rightarrow j}\right]$ in Eq.~\eqref{Expected_value_EADAM}, is computed according to the cumulative function of the Poisson distribution
\begin{equation}
\begin{aligned}
\alpha_{i\rightarrow j} &\equiv 1 - \sum_{x = 0}^{\overline{w}_{i \rightarrow j}^{\mathrm{ts}}-1} P \left(x; \text{E}\left[ \overline{w}_{i\rightarrow j} \right] \right).
\label{p_value_EADAM}
\end{aligned}
\end{equation}
Equation~\eqref{p_value_EADAM} represents the p-value $\alpha_{i \rightarrow j}$, which is used to assess whether the directed pair $i \rightarrow j$ is significant. The same statistical test has to be repeated for directed pairs observed at least once in the overall temporal evolution. For undirected networks, Eq.~\eqref{p_value_EADAM} is equivalent to Eq.~\eqref{p_value_EADM_ETFM}.

\clearpage

\section{Additional simulations}
\label{additional_simulations}

\begin{table}[H]
\centering
\begin{adjustbox}{max width=0.98\textwidth}
\begin{tabular}{c|cc|cc}
  \multirow{4}{*}{Name} & \multicolumn{2}{c|}{Time spent on a DWM}  & \multicolumn{2}{c}{Trading volume exchanged with a DWM}  \\
\cline{2-5}
& \multicolumn{2}{c|}{\thead{Users with stable U2U pairs vs}} & \multicolumn{2}{c}{\thead{Users with stable U2U pairs vs}} \\  
  & \thead{Users with no U2U pairs} &  \thead{Users with any U2U pairs} & \thead{Users with no U2U pairs} &  \thead{Users with any U2U pairs} \\
  & \thead{(KS; p-value)} & \thead{(KS; p-value)} & \thead{(KS; p-value)} & \thead{(KS; p-value)} \\
\hline
\hline 
\thead{Abraxas}	&	\thead{(0.572; 0.0001)}	&	\thead{(0.450; 0.0001)}	&	\thead{(0.382; 0.0001)}	&	\thead{(0.304; 0.0001)}	 \\ 
\thead{Agora}	&	\thead{(0.625; 0.0001)}	&	\thead{(0.522; 0.0001)}	&	\thead{(0.381; 0.0001)}	&	\thead{(0.310; 0.0001)}	 \\ 
\thead{AlphaBay}	&	\thead{(0.583; 0.0001)}	&	\thead{(0.516; 0.0001)}	&	\thead{(0.355; 0.0001)}	&	\thead{(0.280; 0.0001)}	 \\ 
\thead{Apollon}	&	\thead{(0.566; 0.0001)}	&	\thead{(0.487; 0.0001)}	&	\thead{(0.386; 0.0001)}	&	\thead{(0.364; 0.0001)}	 \\ 
\thead{Basetools}	&	\thead{(0.510; 0.0001)}	&	\thead{(0.481; 0.0001)}	&	\thead{(0.390; 0.0001)}	&	\thead{(0.386; 0.0001)}	 \\ 
\thead{Benumb}	&	\thead{(0.527; 0.0001)}	&	\thead{(0.478; 0.0001)}	&	\thead{(0.292; 0.0001)}	&	\thead{(0.263; 0.0001)}	 \\ 
\thead{BitBazaar}	&	\thead{(0.488; 0.0001)}	&	\thead{(0.462; 0.0001)}	&	\thead{(0.373; 0.0001)}	&	\thead{(0.402; 0.0001)}	 \\ 
\thead{Black Bank}	&	\thead{(0.459; 0.0001)}	&	\thead{(0.321; 0.0001)}	&	\thead{(0.287; 0.0001)}	&	\thead{(0.283; 0.0001)}	 \\ 
\thead{Blue Sky}	&	\thead{(0.577; 0.0001)}	&	\thead{(0.488; 0.0001)}	&	\thead{(0.334; 0.0001)}	&	\thead{(0.353; 0.0001)}	 \\ 
\thead{Buybest}	&	\thead{(0.539; 0.0001)}	&	\thead{(0.516; 0.0001)}	&	\thead{(0.334; 0.0001)}	&	\thead{(0.341; 0.0001)}	 \\ 
\thead{Bypass}	&	\thead{(0.590; 0.0001)}	&	\thead{(0.565; 0.0001)}	&	\thead{(0.433; 0.0001)}	&	\thead{(0.407; 0.0001)}	 \\ 
\thead{DarkMarket}	&	\thead{(0.649; 0.0001)}	&	\thead{(0.614; 0.0001)}	&	\thead{(0.442; 0.0001)}	&	\thead{(0.464; 0.0001)}	 \\ 
\thead{Dream}	&	\thead{(0.498; 0.0001)}	&	\thead{(0.416; 0.0001)}	&	\thead{(0.451; 0.0001)}	&	\thead{(0.442; 0.0001)}	 \\ 
\thead{Empire}	&	\thead{(0.502; 0.0001)}	&	\thead{(0.447; 0.0001)}	&	\thead{(0.396; 0.0001)}	&	\thead{(0.379; 0.0001)}	 \\ 
\thead{Evolution}	&	\thead{(0.600; 0.0001)}	&	\thead{(0.485; 0.0001)}	&	\thead{(0.343; 0.0001)}	&	\thead{(0.252; 0.0001)}	 \\ 
\thead{FEshop}	&	\thead{(0.644; 0.0001)}	&	\thead{(0.621; 0.0001)}	&	\thead{(0.446; 0.0001)}	&	\thead{(0.441; 0.0001)}	 \\ 
\thead{Flugsvamp 2.0}	&	\thead{(0.575; 0.0001)}	&	\thead{(0.515; 0.0001)}	&	\thead{(0.444; 0.0001)}	&	\thead{(0.393; 0.0001)}	 \\ 
\thead{Flugsvamp 3.0}	&	\thead{(0.617; 0.0001)}	&	\thead{(0.583; 0.0001)}	&	\thead{(0.508; 0.0001)}	&	\thead{(0.487; 0.0001)}	 \\ 
\thead{FuLLzShOp}	&	\thead{(0.515; 0.0001)}	&	\thead{(0.497; 0.0001)}	&	\thead{(0.207; 0.0001)}	&	\thead{(0.270; 0.0001)}	 \\ 
\thead{Hansa}	&	\thead{(0.583; 0.0001)}	&	\thead{(0.527; 0.0001)}	&	\thead{(0.348; 0.0001)}	&	\thead{(0.366; 0.0001)}	 \\ 
\thead{Hydra}	&	\thead{(0.541; 0.0001)}	&	\thead{(0.530; 0.0001)}	&	\thead{(0.306; 0.0001)}	&	\thead{(0.335; 0.0001)}	 \\ 
\thead{Joker's Stash}	&	\thead{(0.654; 0.0001)}	&	\thead{(0.613; 0.0001)}	&	\thead{(0.399; 0.0001)}	&	\thead{(0.360; 0.0001)}	 \\ 
\thead{LuxSocks.ru}	&	\thead{(0.684; 0.0001)}	&	\thead{(0.599; 0.0001)}	&	\thead{(0.369; 0.0001)}	&	\thead{(0.417; 0.0001)}	 \\ 
\thead{Matanga}	&	\thead{(0.546; 0.0001)}	&	\thead{(0.538; 0.0001)}	&	\thead{(0.329; 0.0001)}	&	\thead{(0.464; 0.0001)}	 \\ 
\thead{Middle Earth}	&	\thead{(0.515; 0.0001)}	&	\thead{(0.368; 0.0001)}	&	\thead{(0.342; 0.0001)}	&	\thead{(0.252; 0.0001)}	 \\ 
\thead{MrGreen.ws}	&	\thead{(0.543; 0.0001)}	&	\thead{(0.514; 0.0001)}	&	\thead{(0.287; 0.0001)}	&	\thead{(0.259; 0.0001)}	 \\ 
\thead{Nightmare}	&	\thead{(0.424; 0.0001)}	&	\thead{(0.350; 0.0001)}	&	\thead{(0.405; 0.0001)}	&	\thead{(0.375; 0.0001)}	 \\ 
\thead{Nucleus}	&	\thead{(0.601; 0.0001)}	&	\thead{(0.477; 0.0001)}	&	\thead{(0.403; 0.0001)}	&	\thead{(0.332; 0.0001)}	 \\ 
\thead{Pandora}	&	\thead{(0.577; 0.0001)}	&	\thead{(0.391; 0.0001)}	&	\thead{(0.334; 0.0001)}	&	\thead{(0.268; 0.0001)}	 \\ 
\thead{Russian Anonymous}	&	\thead{(0.688; 0.0001)}	&	\thead{(0.674; 0.0001)}	&	\thead{(0.532; 0.0001)}	&	\thead{(0.530; 0.0001)}	 \\ 
\thead{San-Wells}	&	\thead{(0.567; 0.0001)}	&	\thead{(0.512; 0.0001)}	&	\thead{(0.310; 0.0001)}	&	\thead{(0.273; 0.0001)}	 \\ 
\thead{Sheep}	&	\thead{(0.528; 0.0001)}	&	\thead{(0.384; 0.0001)}	&	\thead{(0.356; 0.0001)}	&	\thead{(0.316; 0.0001)}	 \\ 
\thead{Silk Road 2}	&	\thead{(0.586; 0.0001)}	&	\thead{(0.462; 0.0001)}	&	\thead{(0.395; 0.0001)}	&	\thead{(0.317; 0.0001)}	 \\ 
\thead{Silk Road 3.1}	&	\thead{(0.566; 0.0001)}	&	\thead{(0.509; 0.0001)}	&	\thead{(0.413; 0.0001)}	&	\thead{(0.412; 0.0001)}	 \\ 
\thead{Silk Road}	&	\thead{(0.598; 0.0001)}	&	\thead{(0.570; 0.0001)}	&	\thead{(0.345; 0.0001)}	&	\thead{(0.331; 0.0001)}	 \\ 
\thead{TradeRoute}	&	\thead{(0.516; 0.0001)}	&	\thead{(0.453; 0.0001)}	&	\thead{(0.377; 0.0001)}	&	\thead{(0.390; 0.0001)}	 \\ 
\thead{Unicc}	&	\thead{(0.714; 0.0001)}	&	\thead{(0.678; 0.0001)}	&	\thead{(0.536; 0.0001)}	&	\thead{(0.521; 0.0001)}	 \\ 
\thead{Valhalla}	&	\thead{(0.473; 0.0001)}	&	\thead{(0.420; 0.0001)}	&	\thead{(0.409; 0.0001)}	&	\thead{(0.478; 0.0001)}	 \\ 
\thead{Wall Street}	&	\thead{(0.569; 0.0001)}	&	\thead{(0.550; 0.0001)}	&	\thead{(0.380; 0.0001)}	&	\thead{(0.418; 0.0001)}	 \\ 
\thead{xDedic}	&	\thead{(0.556; 0.0001)}	&	\thead{(0.457; 0.0001)}	&	\thead{(0.205; 0.0001)}	&	\thead{(0.107; 0.0001)}	 \\ 
\end{tabular}
\end{adjustbox}
\caption{\textbf{Statistical tests.} The two-sided Kolmogorov-Smirnov test is used to perform the statistical test. All p-values are less than 0.0001, which is indicated with 0.0001.}
\label{Statistical_tests}
\end{table}

\clearpage
 
\end{document}